\documentclass[a4paper,fleqn,usenatbib]{mnras}
\pdfoutput=1
\usepackage[T1]{fontenc}
\usepackage{ae,aecompl}

\usepackage{graphicx}	% Including figure files
\usepackage{amsmath}	% Advanced maths commands
\usepackage{amssymb}	% Extra maths symbols
\usepackage{subfig}
\renewcommand{\vec}[1]{\mathbf{#1}}

\title[Magnetic Field Evolution of Neutron Stars]{Magnetic Field Evolution of Neutron Stars I: Basic formalism, numerical techniques, and first results}

\author[A. Bransgrove et al.]{
Ashley Bransgrove,$^{1,2}$\thanks{E-mail: ashley.bransgrove@columbia.edu}
Yuri Levin,$^{1,2,3}$
Andrei Beloborodov$^{2}$
\\
$^{1}$School of Physics and Astronomy, Monash Centre for Astrophysics, Monash University, Clayton, Vic 3800, Australia\\
$^{2}$Physics Department and Columbia Astrophysics Laboratory, Columbia University, 538 West 120th Street, New York, NY 10027\\
$^{3}$Center for Computational Astrophysics, Flatiron Institute, 162 5th Avenue, 6th floor, New York, NY 10010
}

\date{Accepted XXX. Received YYY; in original form ZZZ}

\pubyear{2016}

\begin{document}
\label{firstpage}
\pagerange{\pageref{firstpage}--\pageref{lastpage}}
\maketitle

\begin{abstract}
In this work we explore the evolution of magnetic fields inside strongly magnetized neutron stars in axisymmetry.  We model numerically the coupled field evolution in the core and the crust. Our code models the Hall drift and Ohmic effects in the crust, the back-reaction on the field from magnetically-induced elastic deformation of the crust, the magnetic twist exchange between the crust and the core, and the drift of superconducting flux tubes inside the core. The correct hydromagnetic equilibrium is enforced in the core. We find that i) The Hall attractor found by Gourgouliatos and Cumming in the crust exists also for configurations when the B-field penetrates into the core. However, the evolution timescale for the core-penetrating fields is dramatically different than that of the fields confined to the crust. ii) The combination of Jones' flux tube drift and Ohmic dissipation in the crust can deplete the pulsar magnetic fields 
on the timescale of $150$ Myr if the crust is hot ($T\sim2\times10^8$ K), but acts on much slower timescales for cold neutron stars, such as recycled pulsars ($\sim 1.8 $ Gyr, depending on impurity levels).  iii) The outward motion of superfluid vortices during the rapid spin-down of a young highly magnetized pulsar, can result in a partial expulsion of flux from the core when $B\lesssim 10^{13}$ G. However for $B\gtrsim2\times 10^{13}$ G, the combination of a stronger magnetic field and a longer spin period implies that the core field cannot be expelled.
\end{abstract}


%%%%%%%%%%%%%%%%%%%%%%%%%%%%%%%%%%%%%%%%%%%%%%%%%%

%%%%%%%%%%%%%%%%% BODY OF PAPER %%%%%%%%%%%%%%%%%%

\section{Introduction}

There is a rich variety of neutron stars with magnetic field strengths which differ by several orders of magnitude, and vastly different behaviors. The evolution of neutron star magnetic fields can provide insight into the origin, behavior, and populations of neutron stars [eg. \cite{tauris_understanding_2014}].

Pulsars are rapidly rotating, magnetized neutron stars, and are characterized by intense beams of radio-waves, emitted by the plasma on the open magnetic field lines connecting the star to its light cylinder. Young pulsars have spin periods of tens of milliseconds to seconds, and magnetic field strength of order $10^{11}-10^{12}$ G [see eg. \cite{kaspi_radio_2016}], inferred from dipole braking spin-down. The strongest magnetic field of a radio pulsar is that of PSR J1847-0130, with dipole field magnetic strength $B\simeq9.4\times10^{13}$ G \citep{mclaughlin_psr_2003}. Millisecond pulsars, an extreme subclass of radio pulsars exhibit spin periods ranging from  $1.4$ ms to $\sim10$ ms, and far weaker magnetic fields ranging from $10^8 - 10^{10}$ G. The current fastest spinning millisecond pulsar has spin frequency of $716$ Hz \citep{hessels_radio_2006}. Approximately $60$\% of the $294$ observed pulsars with spin periods less than $10$ ms \citep{manchester_australia_2005} exist in binary systems. This leads to the conclusion that much of the millisecond pulsar population may have been born regular pulsars, which spin-down below the radio death-line by magnetic braking, and then spin-up by the accretion of a companion star. These are the so-called recycled pulsars. The depletion of the pulsar field during this process is not well understood. One proposed mechanism is the burial of the surface field by the accretion flow [see eg. \cite{romani_unified_1990}, \cite{cumming_magnetic_2001}, \cite{choudhuri_diamagnetic_2002}, \cite{payne_burial_2004}]. Another is expulsion of flux from the core, [\cite{ruderman_neutron_1998}, \cite{jones_type_2006}]. It is possible that the origin, birth rate, and spin history of millisecond pulsars can be better understood by studying the evolution of pulsar magnetic fields. 

The most extreme class of neutron stars are magnetars. Canonical magnetars are radio-quiet persistent X-ray sources, powered by dissipation of free energy stored in ultra-strong magnetic fields ($10^{14}-10^{15}$ G) [\cite{thompson_soft_1996}, see also \cite{kaspi_magnetars_2017} for a review]. These young neutron stars have typical spin-down inferred ages of $\tau_c\lesssim 10$ kyr. But the discovery of a wide variety of magnetars over the years has lead to a broader class of objects, with a range of ages and field strengths. Particularly interesting are the so-called transient magnetars, which exhibit extended periods of low luminosity, followed by outbursts, during which the X-ray flux can increase by three orders of magnitude, before returning to quiescence over periods of months to years \citep{turolla_magnetars:_2015}. The mechanism which causes these bursts of activity is not well understood, but is believed to be due to the evolution of super-critical magnetic fields which shear the crust, and generate magnetospheric activity [\cite{perna_unified_2011}, \cite{beloborodov_thermoplastic_2014}, \cite{li_magnetar_2016}, \cite{thompson_global_2017}]. Many transient magnetars are young, seemingly displaying these bursts of activity in the thousands of years following their birth [see eg. \cite{ibrahim_discovery_2004}]. The discovery of SGR 0418+5729 in 2009 with dipole field strength $B=7.5\times10^{12}$ G and a relatively high galactic latitude, was the beginning of a new class of transients significantly older than the commonly accepted magnetar lifetime of $10$ kyr, and with dramatically weaker magnetic fields \citep{rea_low-magnetic-field_2010}. The so-called ``weak-field" magnetars challenge some aspects of the classical magnetar model, which posits that magnetar activity is generated by energy stored in an ultra-strong magnetic field. 

The fact that there are several well defined classes of neutron stars, with a number of transient objects displaying behavior somewhere between these classifications, suggests that the galactic population of neutron stars may be explained by different ages and birth field strengths. The discovery of radio emission from magnetars \citep{camilo_transient_2006}, and X-ray bursts from so-called ``high-B pulsars" \citep{archibald_magnetar-like_2016} adds evidence to this argument. While observations point to a unification of neutron star classes \citep{kaspi_radio_2016}, further theoretical work is required to complete this picture. Recent discussion of a unification was based on the models of magnetic field and thermal evolution \citep{vigano_unifying_2013}. It is likely that any such unification would see a given neutron star traverse a variety of classes over the course of its life, with its classification at any time having a strong dependence on magnetic field strength, and configuration. Indeed studying the evolution of neutron star magnetic fields is key to understanding how a neutron star may transition from one class of object to another.

The evolution of magnetic fields in neutron star crusts is due to Hall drift and Ohmic diffusion. It was studied by \cite{jones_neutron_1988}, and more definitively by \cite{goldreich_magnetic_1992}. Hall drift is the non-linear advection of magnetic fields, by the electron currents supporting $\nabla\times\vec{B}=4\pi/c\vec{j}$. The Hall effect can generate large magnetic shear stresses, countered by the solid stress of the crust. In reality the crust yields elastically to Hall-induced stresses up to a point, beyond which it deforms in the plastic regime [see eg. \cite{levin_dynamics_2012},  \cite{beloborodov_thermoplastic_2014}, \cite{thompson_global_2017}]. Ohmic diffusion is caused by the finite resistivity of the crustal material due to electron scattering by the ion lattice. This process converts magnetic energy to heat, in contrast to Hall drift which conserves magnetic energy. In neutron stars with magnetic fields $B\gtrsim10^{13}$ G, the Hall timescale is shorter than the Ohmic timescale, making Hall drift the dominant channel of evolution for the crustal magnetic field. 

Advances in numerical techniques have allowed Hall drift and Ohmic diffusion of 2D axisymmetric magnetic fields to be studied in numerical simulations, with a variety of pseudo-spectral and grid based methods [\cite{hollerbach_influence_2002}, \cite{pons_magnetic_2007}, \cite{vigano_new_2012}]. The basic finding was that Hall drift could enhance the transfer of magnetic energy to smaller scales, where Ohmic diffusion proceeds more efficiently. \cite{gourgouliatos_hall_2014} found that Hall drift drives itself toward a configuration with uniform electron angular velocity along poloidal field lines. This ``Hall attractor" is analogous to Ferraro's Law in ideal MHD, in which twisted field lines in a cylindrical configuration will evolve to a state with constant angular velocity along field lines \citep{ferraro_non-uniform_1937}. The Hall attractor has interesting implications for the active periods of magnetars and other transients. Most recently Hall drift has been simulated numerically in 3D \citep{wood_three_2015}. \cite{gourgouliatos_magnetic_2016} found that Hall drift can generate localized patches of high magnetic field strength in a magnetar crust, where significant heat can be generated through Ohmic diffusion. 

The majority of numerical studies of magnetic field evolution in neutron star crusts do not include the coupled evolution of the core magnetic field. Considerable work has gone into modeling fields which are confined to the crust, however there is no reason why this should be so. Only one numerical study of the coupled crust-core evolution has been published \citep{elfritz_simulated_2016}. This paper presents our first step in building detailed numerical models for the B-field evolution in neutron stars that include the field in both the crust and the core, with important differences from \cite{elfritz_simulated_2016}. Namely, 1. We enforce the correct hydromagnetic equilibrium in the core, 2. We model the twist exchange between the crust and the core, 3. We consider much faster evolutionary timescales for the core magnetic field, and 4. We model the elastic back-reaction on the magnetic field evolution in the crust.

There are a number of proposed channels of evolution for the core magnetic field. Most well known is the ambipolar diffusion \citep{goldreich_magnetic_1992}, which is the evolution induced by the drift of the charged component through the neutral one, ie. the drift of the proton-electron plasma through the neutrons. Ambipolar diffusion is limited by two factors. Firstly, there is friction between protons and the background neutron fluid. Secondly, departures from chemical equilibrium create pressure gradients which choke the flow of charge currents. Gradients in chemical potential can be erased by weak nuclear interactions. Recently ambipolar diffusion was modeled in 2D by \cite{castillo_magnetic_2017}.

Neutron stars cool as they age and their cores are expected to become superconducting and superfluid (unless the magnetic field is ultra-strong, $B>10^{16}$ G, and quenches superconductivity). This results in the quantization of vorticity into vortex lines and magnetic flux -- into flux tubes. An important magnetic flux transport mechanism is the drift of superconducting flux tubes. \cite{jones_type_2006} shows that flux tubes in a superconducting core can move with viscous dissipation through the core fluid, under their own self tension. The drift of flux tubes has a typical velocity $v\approx4\times10^{-7}$ cm s$^{-1}$ for typical pulsars [see section 3 of \citep{jones_type_2006}], making this effect relevant to the depletion of pulsar magnetic fields. We note straight out that this result is controversial, and there is no consensus about it in the theoretical literature; we discuss it below. Furthermore, \cite{ruderman_rotating_1974} pointed out that the spin-down of superfluid neutron stars must be associated with the outward motion of superfluid neutron vortices. \cite{srinivasan_novel_1990}, and \cite{ruderman_neutron_1998} showed that due to entrainment of superfluid protons the neutron vortices have spontaneous magnetization and that as consequence, there is a strong interaction between superfluid vortices and superconducting flux tubes. In this picture, the flux tubes may be pulled along with neutron vortices during spin-down.

Axisymmetric magnetic configurations satisfy an MHD equilibrium condition in the core, which we formulate and implement in our simulations. We find that with this equilibrium, the magnetic field in the crust and the core asymptotically settles into the Hall Attractor of \cite{gourgouliatos_hall_2014}, which was established for crust-confined fields. We explore the evolution of the core magnetic field under Jones' flux tube drift. Our simulations suggest that a combination of Jones' flux tube drift in the core, and Ohmic diffusion in the crust can deplete pulsar magnetic fields on a timescale of $150$ Myr, if the crust is hot ($T\sim2\times10^{8}$ K). We also consider the hypothesis that a weak-field magnetar can be produced by a neutron star with initially rapid spin and dipole field smaller than the conventional magnetar field. In this scenario, the field is pushed out by the neutron vortices into the crust and thus some of the rotational energy is transformed into magnetic energy. We show that for magnetic fields $B\gtrsim2\times 10^{13}$ G, the combination of a strong magnetic field, and large spin period means the core field cannot be expelled by the vortices. When $B\lesssim 10^{13}$ G, the magnetic field is partially expelled from the core, which launches large scale Hall waves from the crust-core interface. However, these Hall waves are not strong enough to break the crust.

An outline of the paper is as follows. In Section~\ref{formalism} we present the evolution equations for the magnetic field in the crust, for magnetically induced elastic back-reaction of the crust, and for the magnetic field evolution in the core. In Section~\ref{results} we present our results, and in Section~\ref{discussion} we discuss observational implications. Numerical details and derivations of key equations are given in the Appendices.

\section{Equations and Formalism}
\label{formalism}
\subsection{Magnetic Field Evolution in the Crust}\label{crust_evolution}
The evolution of the magnetic field in the crust is governed by the equation
\begin{equation}
\frac{\partial \vec{B}}{\partial t}=\nabla\times\left(\vec{v}\times\vec{B}\right)-\nabla\times\left(\eta\nabla\times\vec{B}\right),
\label{Beveqn1}
\end{equation}
[\cite{jones_neutron_1988}, \cite{goldreich_magnetic_1992}] where the first term here represents advection of the field by the electron fluid with velocity field $\vec{v}$, and the second term represents Ohmic diffusion with diffusivity $\eta=c^2/4\pi\sigma$. Here $\sigma$ is the electrical conductivity of the crust. In contrast to previous work, our model takes into account the velocity of the ion lattice, $\vec{\dot{\xi}}$. The velocity of the electron fluid is then given by
\begin{equation}
\vec{v} = \vec{v}_\text{hall} + \vec{\dot{\xi}},
\end{equation}
where $\vec{v}_\text{hall}$ is the Hall drift velocity,
\begin{equation}
\vec{v}_\text{hall} = -\frac{c}{4\pi n_e e}\nabla\times\vec{B}.
\label{hall_v}
\end{equation}
There are two contributions to $\vec{\dot{\xi}}$,
\begin{equation}
\vec{\dot{\xi}} = \vec{\dot{\xi}}_\text{el} + \vec{\dot{\xi}}_\text{pl}.
\end{equation}
Here $\vec{\dot{\xi}}_\text{el}$ is the elastic deformation, and $\vec{\dot{\xi}}_\text{pl}$ is the plastic deformation. For now we neglect the plastic response of the crust, and content ourselves with the elastic deformation. 

We now follow closely the formalism of \cite{gourgouliatos_hall_2014}. In axisymmetry the magnetic field can decomposed into poloidal and toroidal components, and expressed in terms of the scalar functions $\Psi$ and $I$. The magnetic field is written as a sum of poloidal ($\vec{B}_\text{p}$) and toroidal ($\vec{B}_\text{T}$) components
\begin{equation}
\vec{B} = \vec{B}_\text{p} + \vec{B}_\text{T} = \nabla\Psi\times\nabla\phi + I\nabla\phi,
\label{Bdef}
\end{equation}
where we work in spherical coordinates $(r,\theta,\phi)$, and define $\nabla\phi\equiv \hat{e}_\phi/r\sin\theta$. The function $\Psi$ is known  as the flux function (identical in form to the Stokes stream function), since $2\pi\Psi(r,\theta)$ is the poloidal magnetic flux passing through the polar cap with radius $r$ and opening angle $\theta$. The function $I$ has the interpretation that $cI(r,\theta)/2$ is the poloidal current passing through the same polar cap, and hence is often called the poloidal current function.

We now express the evolution Equation~\eqref{Beveqn1} in terms of the scalar functions $\Psi$ and $I$ as defined in Equation~\eqref{Bdef}. Evolving fields in this formalism has the advantage of automatically preserving $\nabla\cdot\vec{B} = 0$ at all times, provided the scalar functions $\Psi$ and $I$ are differentiable. We begin by defining the quantity
\begin{equation}
\chi= \frac{c}{4\pi en_er^2\text{sin}^2\theta},
\end{equation}
as in \cite{gourgouliatos_hall_2014}. We also write the toroidal current as 
\begin{equation}
\vec{j}_T=\frac{c}{4\pi}\nabla\times\vec{B}_p=-\frac{c}{4\pi}\Delta^{*}\Psi\nabla\phi,
\end{equation}
and the electron angular velocity as 
\begin{equation}
\Omega_\text{e} = \Omega_\text{hall} + \Omega_\text{el} = -\frac{j_T}{r_\bot n_e e} + \frac{v^{\phi}_\text{el}}{r_{\perp}}=\chi\Delta^{*}\Psi + \frac{v^{\phi}_\text{el}}{r_{\perp}},
\label{Omega}
\end{equation}
with $r_\bot\equiv r\sin\theta$ the cylindrical radius, and we have used the Grad-Shafranov operator, 
\begin{equation}
\Delta^{*} = \frac{\partial^2}{\partial r^2}+\frac{\text{sin}\theta}{r^2}\frac{\partial}{\partial\theta} \left( \frac{1}{\text{sin}\theta}\frac{\partial}{\partial\theta}\right).
\end{equation}
Using the above definitions, the Hall evolution equation reduces to the following two scalar equations, in terms of the poloidal and toroidal scalar functions
\begin{equation}
\frac{\partial\Psi}{\partial t} - r^2\text{sin}^2\theta\chi(\nabla I\times\nabla\phi)\cdot\nabla\Psi = \frac{c^2}{4\pi\sigma}\Delta^{*}\Psi,
\label{hall_pol}
\end{equation}
\begin{equation}
\begin{split}
\frac{\partial I}{\partial t} + r^2\text{sin}^2\theta[(\nabla\Omega_\text{e}\times\nabla&\phi)\cdot\nabla\Psi+I(\nabla\chi\times\nabla\phi)\cdot\nabla I]
\\&
 = \frac{c^2}{4\pi\sigma}\left(\Delta^{*}I-\frac{1}{\sigma}\nabla I\cdot\nabla\sigma\right).
 \end{split}
 \label{hall_tor}
\end{equation}
These evolution equations are the same as those in \cite{gourgouliatos_hall_2014}, except for the addition of the elastic back-reaction velocity of the crust. We assume a neutron star radius of $r_*=10$ km, and a crust thickness of $1$ km, so that the crust-core interface is at radius $r_c=9$ km. We use the electron density profile and electrical conductivity provided by \cite{gourgouliatos_hall_2014} who take $n_e\propto z^4$, with $z$ the depth into the crust, and $\sigma\propto n_e^{2/3}$, which is somewhere between the density scalings expected for phonon scattering and impurity scattering. The electron number density at the base of the crust is given as $n_e= 2.5\times10^{36}\text{cm}^{-3}$. We do not include the upper crust with density $\rho<10^{11}$ g cm$^{-3}$ to avoid time step issues in the low density regions where the evolution is very fast. The electrical conductivity varies from $\sigma=3.6\times 10^{24}$ $\text{s}^{-1}$ at the base of the crust, to $\sigma = 1.8\times 10^{23}$ $\text{s}^{-1}$ at the surface, which is appropriate for phonon scattering at $T\approx2\times10^8$~K \citep{gourgouliatos_hall_2014}. We can determine the characteristic timescales of evolution of the Hall and Ohmic terms, with $L$ a characteristic length scale, taken to be the thickness of the crust (1 km) and $n_e$ and $\sigma$ evaluated at the base of the crust,
\begin{equation}
t_{\text{ohm}}\sim\frac{4\pi\sigma L^2}{c^2}=13.5\left(\frac{L}{1\text{km}}\right)^2\left(\frac{\sigma}{3.6\times 10^{24}\text{ s}^{-1}}\right)\text{Myr},
\label{ohmic_time}
\end{equation}
\begin{equation}
t_{\text{hall}}\sim\frac{4\pi eL^2n_e}{cB}=\frac{1.6}{B_{14}}\left(\frac{L}{1\text{km}}\right)^2\left(\frac{n_e}{2.5\times 10^{36}\text{ cm}^{-3}}\right)\text{Myr}
\label{hall_time}
\end{equation}
where $B_{14}$ is the magnetic field strength in units of $10^{14}$G.

\subsection{Matching the Crust Field to the Magnetosphere}
In order to solve the evolution Equations \eqref{hall_pol} and \eqref{hall_tor} we need two boundary conditions, for $I$ and $\Psi$ at $r_*$, which may also be formulated as conditions on $\vec{B}(r_*,\theta)$. In this work we are considering a slow evolution of magnetic fields on timescales over which any episodic magnetospheric twists (magnetar activity) must be erased \citep{beloborodov_untwisting_2009}. It is therefore reasonable to assume a vacuum magnetic field as the boundary condition at the surface, as was previously shown by \cite{marchant_revisiting_2011}, \cite{marchant_stability_2014}, and \cite{castillo_magnetic_2017}. Demanding zero current means that $\nabla\times\vec{B} = 0$ outside the star. So we can write the vacuum field as 
\begin{equation}
\vec{B}=\nabla V,
\label{potential_B}
\end{equation}
where $V$ is a scalar function. We also assume that there is no surface current at $r_*$ due to the finite electrical conductivity of the outer crust. Thus, the two boundary conditions express the continuity of the tangential components of the magnetic field $B_\phi$, $B_\theta$, at the surface, so that they match a vacuum solution outside the star:\\

\noindent
(i) The continuity of $B_\phi$ implies for the crustal field $B_\phi(r_*,\theta)=0$, since in any axisymmetric vacuum magnetosphere  $B_\phi=(r\sin\theta)^{-1}\partial V/\partial \phi=0$. This gives $I(r_*,\theta)=0$.
\\

\noindent
(ii) The continuity of $B_\theta$ gives a condition on $\partial\Psi/\partial r=-r\sin\theta B_\theta$ --- This boundary condition is formulated below (no boundary condition is imposed on the values of $\Psi(r_*,\theta)$ -- its evolution is calculated from Equation~\eqref{hall_pol} in the crust).
\\

The constraint $\nabla\cdot\vec{B}=0$, gives the Laplace equation for $V$, $\nabla^2V=0$. $B_\theta$ is determined by this Laplace equation outside the star for given surface values of $B_r$. Because $V\longrightarrow 0$ as $r\longrightarrow\infty$, we can write the solution as a multipolar expansion. For axisymmetric magnetic fields $V$ is given by
\begin{equation}
V(r,\theta) = \sum_{l=1}^{\infty} \frac{a_{l}}{r^{l+1}} P_l(\cos\theta),
\end{equation}
where $P_l(\cos\theta)$ are the Legendre polynomials. The sum starts at $l=1$ because there are no magnetic monopoles, and $a_{l}$ are expansion coefficients to be determined. Now using the definition of the magnetic field in Equation~\eqref{potential_B}, we can write
\begin{equation}
B_r(r,\theta) = \frac{\partial V}{\partial r} = -\sum_{l=1}^{\infty} (l+1)\frac{a_{l}}{r^{l+2}} P_l(\cos\theta).
\label{Br}
\end{equation}
We can invert this expression to solve for the expansion coefficients, and evaluate it at $r_*$ which gives
\begin{equation}
a_{l} = -r_{*}^{l+2}\frac{2l+1}{2l+2}\int_0^{\pi} P_l(\cos\theta)B_r(r_{*},\theta)\sin\theta d\theta.
\end{equation}
Using these expansion coefficients, we can calculate $B_{\theta}$ using Equation~\eqref{Bdef},
\begin{equation}
B_{\theta}(r,\theta) = \frac{1}{r}\frac{\partial V}{\partial \theta} = \sum_{l=1}^\infty\frac{a_l}{r^{l+2}}\frac{d P_l(\cos\theta)}{d\theta}=-\frac{1}{r\sin\theta}\frac{\partial\Psi}{\partial r}.
\end{equation} 
This gives the boundary condition for $\Psi$ on the surface of the star as
\begin{equation}
\frac{\partial\Psi}{\partial r}\bigg|_{r_*} =-\sin\theta \sum_{l=1}^\infty\frac{a_l}{r_*^{l+1}}\frac{d P_l(\cos\theta)}{d\theta}.
\end{equation}

\subsection{Magneto-elastic Evolution of the Crust}\label{elastic_method}
The elastic response of the neutron star crust to magnetic stress is governed by the elastodynamic wave equation, with an external Lorentz driving force. We begin by defining the Lagrangian displacement field of the neutron star crust as 
\begin{equation}
\vec{\xi}(\vec{r},t)\equiv \vec{r'} - \vec{r},
\end{equation}
where $\vec{r}$ is the position of a point in the crust before deformation, and $\vec{r'}$ is the position of that point after the deformation. The elastodynamic wave equation can be derived from Newton's second law. We restrict ourselves to the regime of linear elastodynamics and only consider small, reversible deformations of the crust. In this framework the stress depends linearly on the displacement, and we may express the stress as the sum of elastic stress tensor using Hooke's Law, and the magnetic stress using the magnetic part of the Maxwell tensor. It has been shown that the speed at which shear waves propagate in a neutron star crust is remarkably constant over its depth \citep{ruderman_crystallization_1968}. We work in the approximation that shear waves propagate at constant speed [$v_\text{sh}\approx10^{8}\text{ cm s}^{-1}$, \cite{ruderman_crystallization_1968}, \cite{mcdermott_nonradial_1988}] throughout the crust, and also that they propagate at the same speed in all directions (isotropic). We estimate the shear modulus as
\begin{equation}
\mu = \rho v_\text{sh}^2 \approx 10^{28}\rho_{12}\text{ erg cm}^{-3},
\end{equation}
where we assume the mass density scaling $\rho\propto z^{8}$. Specifically, we chose $\rho_{12} = 0.5[(1.1r_* - r)/0.1r_*]^8$, which varies from $\rho=1.3\times10^{14}$ g cm$^{-3}$ at the base of the crust, to $\rho=5\times 10^{11}$ g cm$^{-3}$, our chosen surface cutoff. Additionally, assuming an incompressible crust ($\nabla\cdot\vec{\xi}=0$), and a spherically symmetric unevolving shear modulus $\mu(r)$ yields the elastodynamic wave equation
\begin{equation}
\rho\frac{\partial^2\vec{\xi}}{\partial t^2} = (\nabla\mu \cdot\nabla)\vec{\xi} - (\vec{\xi}\cdot \nabla)\nabla\mu  + \mu\nabla^2\vec{\xi} + \frac{1}{4\pi}(\nabla\times\vec{B})\times\vec{B}.
\label{elastic}
\end{equation}
The term on the left-hand side (LHS) of Equation~\eqref{elastic} is due to the inertia of the crust. Since Hall drift occurs on timescales much longer than the elastic wave crossing time, we may neglect the inertial term in these models. The first three terms on the right-hand side (RHS) of Equation~\eqref{elastic} are due to the elastic restoring forces of the solid crust, and the last term is due to Maxwell stresses. With the neglect of the inertial term, magneto-elastic equilibrium is given by 
\begin{equation}
\frac{1}{4\pi}(\nabla\times\vec{B})\times\vec{B} = -(\nabla\mu \cdot\nabla)\vec{\xi} + (\vec{\xi}\cdot \nabla)\nabla\mu  - \mu\nabla^2\vec{\xi}.
\end{equation}
We, however, prefer to deal with an evolution (rather than the above constraint) equation for $\xi$ and therefore introduce a small non-zero velocity
\begin{equation}
\vec{v}_\text{el}\equiv\frac{\partial\vec{\xi}}{\partial t}.
\end{equation}
It corresponds to a small deviation from the force balance that we write in a relaxation/damping form $\vec{f}_\text{damp}=-\gamma\rho \vec{v}_\text{el}$. The value of $\gamma$ is not important as long as it is small enough, so that the system evolves very close to the force balance. Effectively, this is a dynamic way of implementing the constraint on $\xi$ required by the force balance. This yields the evolution equation,
\begin{equation}
\begin{split}
\frac{\partial\vec{\xi}}{\partial t}& = \frac{1}{\gamma\rho}\left[(\nabla\mu \cdot\nabla)\vec{\xi} - (\vec{\xi}\cdot \nabla)\nabla\mu  + \mu\nabla^2\vec{\xi}\right]
\\&
+ \frac{1}{4\pi\gamma\rho}(\nabla\times\vec{B})\times\vec{B},
\end{split}
\label{elasto_mag}
\end{equation}
which when evolved in the limit of small $\gamma$ will tend toward the adiabatic solution. This relaxation method is equivalent to solving a matrix problem to find $\xi$, but avoids the difficulty of equations which are implicit in the evolution of $\vec{B}$. The challenge in choosing the value of $\gamma$ is on the one hand to ensure the relaxation is fast enough so that the crust is in equilibrium between the magnetic and elastic forces, but on the other hand is slow enough so that the numerical computations do not become too costly. It is helpful to consider the characteristic relaxation timescale
\begin{equation}
t_\text{re} = \gamma\frac{ L^2}{v_\text{sh}^2},
\end{equation}
and require $t_\text{re}\ll t_\text{hall}$. This gives the criterion
\begin{equation}
\gamma \ll \frac{4\pi n_e e}{B} v_\text{sh}^2. 
\end{equation}
The back reaction of the crustal motion on the evolution of the magnetic field occurs through the equation
\begin{equation}
\frac{\partial\vec{B}}{\partial t} = \nabla\times\left[ (\vec{v}_\text{hall} + \vec{v}_\text{el})\times \vec{B}\right] + \nabla\times(\eta\nabla\times\vec{B}),
\end{equation}
as shown in the previous section.

\subsection{Magnetic Field Evolution in the Core}
There are a number of proposed mechanisms which can transport magnetic flux in neutron star cores, and may lead to an understanding of numerous observable phenomena. Below we outline each of the effects we implement in our numerical scheme, motivate the equations of motion, and discuss the relevant timescales of evolution. We assume a simplified model for the core consisting of protons, neutrons and electrons. We also assume that the neutrons are superfluid, and the protons exist in a type-II superconducting state \citep{baym_superfluidity_1969}.
Observations of the Cas A remnant indicate that the core is likely to be superfluid and superconducting $\approx 300$ yr after birth, and at a temperature $7-9\times10^{8} $ K (greater than the temperatures we consider in our models) [\cite{shternin_cooling_2011}, \cite{page_rapid_2011}]. Hence, the physics of the phase transition itself may be safely neglected in this work.

\subsubsection{Hydromagnetic equilibrium} \label{hydromagnetic_theory}
In this section we outline two methods for studying the evolution of neutron star core magnetic fields on long (Hall) timescales, while maintaining stability on dynamical (MHD) timescales, and taking into account fluid degrees of freedom. The core has a very high conductivity and we treat it as an ideal conductor, so that the field is perfectly coupled to the fluid. It is then instructive to consider displacements of the fluid, since these correspond directly to degrees of freedom of the field. Firstly we assume there is no bulk fluid displacement in the radial direction $\xi_r=0$ on the Alfv\'{e}n timescale, and that the core is incompressible, which in axisymmetry implies $\xi_\theta=0$. In this axisymmetric model, the core fluid can only be displaced in the azimuthal direction, which corresponds to the motion of fluid elements on spherical shells, at fixed cylindrical radius. Such displacements in the $\phi$ direction do not perturb the local pressure or chemical potential, and are only limited by the viscosity of the fluid which is negligible. Thus, it is a good approximation to assume that any toroidal flux injected into the core, readily distributes itself according to a tension equilibrium along poloidal field lines.  In axisymmetry this corresponds to $f_\phi = \vec{j}_p\times\vec{B}_p /c = 0$, but in general it is the vanishing of the solenoidal part of the Lorentz force. In terms of the scalar functions $I$ and $\Psi$, this condition is equivalent to
\begin{equation}
I = I(\Psi),
\end{equation}
which means that $I$ is constant along poloidal field lines in the core. We present two methods enforcing this condition. We briefly outline the first method here, leaving the details for Appendix \ref{AppendixA}. The second method is more general, and we discuss it in more detail.

Firstly, it is possible to determine the value of $I$ along a given poloidal field line in the core by calculating the advection flux of $B_\phi$ into the core by Hall drift. The advection flux of $B_\phi$ is defined by writing the Hall evolution equation for $B_\phi$ in the crust in conservative form, and identifying the advection flux. In this method it is convenient to work in the so-called flux-coordinates $(\Psi,\lambda,\phi)$, where $\Psi$ labels surfaces of constant poloidal flux, and $\lambda$ is the length along a given poloidal field line in the $\phi=const$ plane [e.g. \cite{goedbloed_advanced_2010}]. It can be shown (Appendix \ref{AppendixA}), that the twist angle $\zeta$ of a given poloidal field line in the core evolves according to the equation
\begin{equation}
\frac{\partial\zeta(\Psi)}{\partial t} = -[J(\Psi,\lambda_2,t) - J(\Psi,\lambda_1,t)],
\label{twistev}
\end{equation}
where we have identified the twist angle 
\begin{equation}
\zeta(\Psi) = \int_{\lambda_1}^{\lambda_2} d\lambda \left(\frac{B_\phi}{r_\bot B_\lambda}\right),
\label{twistint}
\end{equation}
with the integral taken along the magnetic field line ($\Psi=\text{const}$). $J$ is related to the ``flux of twist" into the core through
\begin{equation}
F = r_\bot B_\lambda J = v_\lambda B_\phi - v_\phi B_\lambda.
\end{equation}
The RHS of Equation~\eqref{twistev} represents the difference between flux of twist at each footpoint of a field line threading the core at its boundary. There are two contributions to the flux of twist $J$. The first term can be attributed to Hall drift advecting $B_\phi$ into the core with poloidal drift currents, and the second term is the azimuthal winding of poloidal field lines by Hall drift. Equation~\eqref{twistev} may be rearranged to obtain the following equation for the evolution of $I(\Psi)$,
\begin{equation}
\frac{ \partial I(\Psi,t)}{\partial t}= -\varpi(\Psi)[J(\Psi,\lambda_2,t) - J(\Psi,\lambda_1,t)],
\label{coreev}
\end{equation}
with
\begin{equation}
\varpi(\Psi) = \left(\int_{\lambda_1}^{\lambda_2} \frac{d\lambda}{r_\bot^2 B_\lambda}\right)^{-1}.
\label{integral}
\end{equation}
If the toroidal field displays equatorial plane reflection symmetry, and the poloidal field displays equatorial symmetry, $J(\Psi,\lambda_1,t) = J(\Psi,\lambda_2,t)$,
and there will be no magnetic twist injected into the core. 

The procedure outlined above is efficient in tracing the crust-core evolution of the field, so long as the poloidal magnetic field lines in the core are fixed in time. However, in our studies we would like to have the freedom to evolve the core poloidal field. In such situations, it is more practical to use the second method outlined below.

The second method of enforcing hydromagnetic equilibrium, is to treat it as a relaxation problem. This method has the advantage of not requiring the integral in Equation~\eqref{integral} to be evaluated. The principle of the method is similar to that used in Section~\ref{elastic_method} for computing the elastic response of the crust. Suppose there is a poloidal field threading the core. Hall drift in the crust will slowly displace the magnetic field lines in the azimuthal direction, and in response to this, the core field will adjust, quickly returning to hydromagnetic equilibrium. The evolution of the core field can be written
\begin{equation}
\frac{\partial\vec{B_\text{T}}}{\partial t} = \nabla\times (\vec{v}_\text{T}\times\vec{B}_\text{p} + \vec{v}_\text{p}\times\vec{B}_\text{T}),
\label{relaxation}
\end{equation}
with $\vec{v}_\text{T}$ the toroidal (azimuthal) velocity which returns the core field to hydromagnetic equilibrium. The second term on the RHS is the advection of the toroidal field with the poloidal drift velocity $\vec{v}_\text{p}$, of flux surfaces. This term ensures that each poloidal field line maintains its own twist angle when the flux surfaces are evolving. For a given (fixed) poloidal field configuration, equilibrium is satisfied when
\begin{equation}
\frac{\partial\vec{B_\text{T}}}{\partial t} =0 \quad \text{ (equilibrium)}.
\end{equation}
Thus all that is required, is to choose a convenient form of $\vec{v}_\text{T}$ which drives the field towards equilibrium faster than the other channels of evolution, such as Hall drift. First though, we write Equation \eqref{relaxation} in a more convenient form. It can be shown that Equation \eqref{relaxation} may be written as
\begin{equation}
\frac{\partial B_\phi}{\partial t} + \nabla_\text{p}\cdot(\vec{v}_\text{p}B_\phi) = \nabla_\text{p}\cdot(v_\phi\vec{B}_\text{p}),
\label{toroidal_cty}
\end{equation} 
where we have defined the poloidal differential operator
\begin{equation}
\nabla_\text{p}\equiv \left( \frac{\partial}{\partial r_\bot}, \frac{
\partial}{\partial z}\right),
\end{equation}
which acts in the 2D plane. Equation~\eqref{toroidal_cty} is easily interpreted as a continuity equation, with the second term on the LHS the divergence of a transport flux of $B_\phi$ due to drift of the poloidal field. The term on the RHS is a source term, which injects or extracts $B_\phi$ (and magnetic twist) from the magnetic field lines which enter the crust. When using this method $\vec{v}_\text{p}$ must be the same as the poloidal drift velocity of flux surfaces in the core. While in this paper we only use this method for the case of fixed poloidal field lines in the core ($\vec{v}_\text{p}=0$), the method is also applicable to the case of non-zero $\vec{v}_\text{p}$, which we discuss later. For the simplified case of $\vec{v}_\text{p}=0$ though, the second term on the LHS of Equation~\eqref{toroidal_cty} vanishes. Then, all that remains is to choose a convenient form of $v_\phi$ which will drive the field toward equilibrium. We find that such a form is 
\begin{equation}
v_\phi = \frac{k}{|\vec{B}_\text{p}|}(\nabla I\cdot\hat{\vec{e}}_{\lambda}),
\label{hydro_v}
\end{equation}
which obviously tends to zero as the field is driven toward equilibrium. Here $k$ is a relaxation parameter to be be tuned. For the case of static poloidal fields in the core ($\vec{v}_\text{p}=0$), Equation~\eqref{toroidal_cty} becomes
\begin{equation}
\frac{\partial I}{\partial t} = r_\perp \nabla_\text{p}\cdot[k(\nabla I\cdot\hat{\vec{e}}_{\lambda}) \hat{\vec{e}}_{\lambda}],
\label{diffusion_relax}
\end{equation}
which has the form of a modified diffusion equation. It is convenient to work with this form of the equation, since diffusion equations are less problematic to solve numerically. Equation~\eqref{diffusion_relax} with a large $k$ ensures that the magnetic field in the core evolves through a sequence of MHD equilibria, and that these equilibria are stable.

In the more general case, poloidal magnetic field lines are not fixed in the core, but can drift with Ambipolar diffusion, or Jones drift for example. In this case one must ensure that $\vec{v}_\text{p}$ matches the drift velocity of field lines in the crust at the crust-core interface. In the crust, the velocity of poloidal field lines  is due to the Hall drift velocity $\vec{v}_\text{hall}$, and the Ohmic diffusion velocity $\vec{v}_\text{ohm}$,
\begin{equation}
\vec{v}_\text{p}(r_c) = \vec{v}_\text{hall}(r_c) + \vec{v}_\text{ohm}(r_c),
\end{equation}
where $\vec{v}_\text{hall}$ is determined by Equation \eqref{hall_v}. We determine the Ohmic drift velocity of poloidal field lines by noting that the electric field determines $\partial \vec{B}/\partial t = -c \nabla \times \vec{E}$. If we assume that  $\vec{v}$ is perpendicular to $\vec{B}$, one may rewrite the evolution equation as $\partial \vec{B}/\partial t=\nabla \times (\vec{v} \times \vec{B})$ with $\vec{v}$ defined by $\vec{E} = -\vec{v} \times \vec{B}/c$. Using Ohm's law, for the case of poloidal fields this becomes
\begin{equation}
\frac{\vec{J}_\text{T}}{\sigma} = \frac{c}{4\pi\sigma}\nabla\times\vec{B}_\text{p} = -\frac{1}{c}\vec{v}_\text{p}\times\vec{B}_\text{p},
\end{equation}
with $\vec{v}_\text{p}$ the poloidal velocity. Taking the cross product of both sides with $\vec{B}_\text{p}$ allows us to solve for $\vec{v}_\text{p}$. This is the velocity at which poloidal magnetic field lines drift due to Ohmic diffusion, and we call it $\vec{v}_\text{ohm}$,
\begin{equation}
\vec{v}_\text{ohm} = \frac{c^2}{4\pi\sigma}\frac{[(\nabla\times\vec{B}_\text{p})\times\vec{B}_\text{p}]}{(\vec{B}_\text{p}\cdot\vec{B}_\text{p})}.
\end{equation} 
When this velocity is inserted into an induction equation it is exactly equivalent to the Ohmic diffusion equation, so this is indeed the correct Ohmic diffusion velocity. 

We note here that a number of previous works fail to include the correct hydromagnetic equilibrium in the core, rendering their boundary condition on the crust-core interface unphysical. \cite{suvorov_gravitational_2016} violate equatorial plane reflection symmetry, and therefore effectively must be injecting magnetic twist into the core within the timescale of their simulation. The simulations of \cite{elfritz_simulated_2016} include strong toroidal fields in the core, which in general do not satisfy $f_\phi=0$. These stresses cannot be supported by the fluid, and therefore these simulations violate hydromagnetic equilibrium. This error in \cite{elfritz_simulated_2016} is due to their lack of the terms which advect magnetic field by an azimuthal fluid motion; in other words, the background in which their flux tubes move is assumed to be static.

\subsubsection{Jones flux tube drift}\label{Jones_theory}
In newborn neutron stars the magnetic field undergoes a period of dynamically unstable evolution before settling into a stable configuration in which the stresses of the field, given by the Maxwell stress tensor, are balanced by the fluid degrees of freedom. Many Alfv\'{e}n crossing times later, the crust solidifies, and cooling of the core below $T_\text{crit}\approx10^8 - 10^9$ K is accompanied by Cooper pairing of protons to form a $^1S_0$ superfluid [eg., \cite{baym_neutron_1975}].  The phase transition to superconductivity is associated with the quantization of magnetic flux on microscopic scales, with the quantum of flux $\phi_0=hc/2e$. The flux is localised within proton supercurrent vortices, and drops off exponentially on the penetration-depth scale $\lambda\lesssim10^{-11}$ cm. The mean intervortex spacing is $d=5\times10^{-10}$ $B_{12}^{-1/2}$cm, greater than the penetration depth $\lambda$, so that the flux tubes are very weakly interacting. \cite{jones_alignment_1975-1} realized that the anisotropic component of the magnetic stress tensor in type-II superconductors can be significantly larger than $B^2/4\pi$. \cite{easson_stress_1977} showed that in the limit of $B<H_{c1}$, with $H_{c1}\approx10^{15}$ G the lower critical limit, the stress tensor is given as
\begin{equation}
\sigma_{ij} = P_{\text{matter}}\delta_{ij} + \frac{H_i B_j}{4\pi},
\label{tensor}
\end{equation}
where $H_i$ are the components of vector ${\mathbf H}$. Matter contributes the isotropic component of $\sigma_{ij}$, and accounts for the buoyancy of flux tubes; it will be neglected below. As expected the magnetic contribution to the isotropic component is suppressed, due to the lack of magnetic pressure between neighbouring flux tubes. Taking the divergence of this stress tensor gives the volume force
\begin{equation}
\vec{f}_B = \frac{1}{4\pi}(\vec{B}\cdot\nabla)\vec{H}_\text{c1},
\label{tension_force}
\end{equation}
where we have assumed  $H\approx{H}_{c1}$. Note that $\vec{H}_{c1}$ and $\vec{B}$ are locally parallel vectors, and $\vec{B}$ is the spatial average of the microscopic magnetic field. The flux tubes drift with velocity $\vec{v}_L$ through the core fluid, driven by the tension force of the flux localised within the tubes $\vec{f}_B$. A simple estimate of the buoyancy force discussed by \cite{dommes_vortex_2017} indicates that it is often smaller than the tension force Equation~\eqref{tension_force} for typical configurations. Therefore, in this work we do not consider magnetic buoyancy, and neglect the term $P_\text{matter}\delta_{ij}$ in Equation~\eqref{tensor}. The drift velocity is defined by considering the balance of forces on the fluid in the vicinity of a flux tube. We follow \cite{jones_type_2006}, who writes the force balance as
\begin{equation}
\begin{split}
\vec{f}_B - \frac{(n_e e)^2}{\tilde{\sigma}}(\vec{v}_L - &\vec{v}_e) - \frac{n_e e}{c}(\vec{v}_L - \vec{v}_e)\times\vec{B}
\\&
- \frac{n_p e}{c}(\vec{v}_{p0} - \vec{v}_L)\times\vec{B} = 0,
\end{split}
\label{force_balance}
\end{equation}
with $n_e$ and $n_p$ the electron and proton density, $\tilde{\sigma}$ an effective electron conductivity in the superconductor\footnote{Note that $\tilde{\sigma}$ has the same mathematical form as $\sigma$ [e.g. \cite{baym_superfluidity_1969}]. The key difference as noted by \cite{jones_neutron_1991}, is that the electron transport relaxation time is modified when the scattering sites are localised within flux tubes [confusingly, \cite{jones_neutron_1991} denotes $\tilde{\sigma}$ by $\sigma$]. Therefore $\tilde{\sigma}$ is a different quantity to $\sigma$ in normal matter.} and $\vec{v}_{e}$ and $\vec{v}_{p0}$ the macroscopic electron and proton drift velocities respectively. \cite{jones_neutron_1991} derives each of the terms in Equation~\eqref{force_balance}. The second term on the LHS of Equation~\eqref{force_balance} is due to drag from scattering of electrons on quasi-particles localised in the flux tube core, the third term is the Lorentz force on the electrons, and the fourth term is the Magnus force, due to motion of the flux tube through the superfluid protons. 

In a multifluid type-II superconductor, macroscopic currents due to relative motion between charged species are suppressed, and the electrons and protons are co-moving on large scales. This can be expressed with the charge current screening condition of \cite{jones_neutron_1991},
\begin{equation}
\vec{J}^e + \vec{J}^p = 0,
\end{equation}
where $\vec{J}^e$ is the electron current density, and $\vec{J}^{p}$ the proton supercurrent density. This ensures satisfaction of Amperes law, and the London equation throughout the superconductor. For the purposes of this work we neglect entrainment, and using $\vec{J}^e = -n_e e \vec{v}_e$, and $\vec{J}^p = n_p e \vec{v}_{p0}$, it is easy to see that 
\begin{equation}
n_e \vec{v}_e = n_p \vec{v}_{p0}. 
\end{equation}
Implementing this screening condition, and assuming charge neutrality ($n_e=n_p$) leads to cancellation of the Magnus and Lorentz forces, and the drift velocity of flux tubes is given by the expression
\begin{equation}
\vec{v}_L = \vec{v}_e + \frac{\tilde{\sigma}}{n_e^2e^2}\vec{f}_B.
\label{drift_velocity}
\end{equation}
The flux tube velocity has two contributions -- the motion of flux tubes with the charged fluid, and the motion of flux tubes through the charged fluid. Here $\vec{v}_e$, represents motion of flux tubes with the charged plasma (what is usually referred to as ambipolar diffusion). Motion of the flux tubes faster than the background charged plasma occurs with the Jones velocity 
\begin{equation}
\vec{v}_\text{J} = \frac{\alpha}{4\pi}(\vec{B}\cdot\nabla)\vec{H}_\text{c1},\quad \alpha=\tilde{\sigma}/n_e^2e^2,
\label{jones_v}
\end{equation}
and is accompanied by dissipation at the rate $\vec{v}_\text{J}\cdot\vec{f}_B$ -- here $\alpha$ is an effective drag coefficient. It is important to note that the transport velocity of flux can be significantly larger than the plasma drift velocity (ambipolar velocity), which tends to be slowed by the formation of sharp pressure gradients in typical cases. In this work we neglect the velocity $\vec{v}_e$ (ambipolar diffusion), and consider only the Jones drift velocity (ie. we set $\vec{v}_L = \vec{v}_\text{J}$). It was recently shown by \cite{passamonti_relevance_2017} that ambipolar diffusion in superfluid neutron stars can be significant for solenoidal flows at low temperatures ($T<10^{9}$~K). The detailed consideration of ambipolar diffusion is outside of the scope of this work.

It is important to note that this drift velocity is very different to that of \cite{glampedakis_magnetohydrodynamics_2011}, \cite{graber_magnetic_2015}, and \cite{elfritz_simulated_2016}. This can be traced to differences in the calculation of forces acting on a flux tube by \cite{glampedakis_magnetohydrodynamics_2011} (private communication). Clarifying this difference requires a separate investigation and is outside of the scope of this work. It should be noted that the force balance \eqref{force_balance} is non-trivial, and careful calculation is required to determine these forces [see \cite{nozieres_motion_1966} and \cite{jones_neutron_1991}]. 

In a type-II superconductor the motion of flux tubes with velocity $\vec{v}_L$ induces an electric field localized within the vortex cores [see eg. \cite{nozieres_motion_1966}, \cite{parks_superconductivity:_1969}, \cite{jones_type_2006}]. The spatial average of this electric field is 
\begin{equation}
\vec{E} = -\frac{1}{c}\vec{v}_L\times\vec{B},
\end{equation}
which upon substitution into Faraday's law, leads to the following evolution equation for the poloidal part of the spatially averaged field,
\begin{equation}
\frac{\partial \vec{B}_\text{p}}{\partial t} = \nabla\times(\vec{v}_L\times\vec{B}_\text{p}),
\label{pol_ev_eqn}
\end{equation}
while the evolution of the toroidal field on comparable timescales is such that it satisfies hydromagnetic equilibrium at all times. Equation~\eqref{pol_ev_eqn} is very different from the evolution equation used by \cite{elfritz_simulated_2016} (see their equation 16). Firstly, as pointed out by \cite{dommes_vortex_2017}, the magnetic field is locked into flux tubes, and therefore, the evolution of the magnetic field must be governed by an advection equation of the form Equation~\eqref{pol_ev_eqn}, where the field is advected at the same velocity as the flux tubes, $\vec{v}_L$. However the evolution equation of \cite{elfritz_simulated_2016} does not advect the magnetic field at the velocity $\vec{v}_L$.

As in Section~\ref{crust_evolution} we write the evolution equation in terms of the scalar functions $\Psi$ and $I$,
\begin{equation}
\frac{\partial\Psi}{\partial t} + \vec{v}_\text{J}\cdot\nabla\Psi=0,
\label{jones_eqn}
\end{equation}
while $I$ satisfies Equation \eqref{coreev}. \cite{jones_type_2006} estimates $\tilde{\sigma}\approx 10^{29} - 10^{32} B_{12}^{-1}$ s$^{-1}$, depending on the composition of the core. We use in our simulations $\tilde{\sigma}=10^{29}B_{12}^{-1}\text{ s}^{-1}$. For a $1.4M_\odot$ neutron star, a typical baryon density at the centre of the core is $n_B \approx 3.5 \times 10^{38}$ cm$^{-3}$ \citep{li_numerically_2016}. We take a central electron fraction $Y_e=0.1$, which gives $n_e = Y_e n_B = 3.5 \times 10^{37}$ cm$^{-3}$. Rather than adopt a particular equation of state we use these conservative values to calculate $\alpha$ throughout the core, which will cause the field evolution to be slower in the outer core in our simulations. But for our purposes we want to understand the dynamics of flux tubes on long timescales, and this will not affect the end state of our simulations.  

\subsubsection{Flux transport by neutron vortices}\label{vortex_theory}
\begin{table*}
\caption{Summary of the models we discuss in Section~\ref{results}. Here HME (relaxation) implies hydromagnetic equilibrium enforced by the relaxation method in Section~\ref{hydromagnetic_theory}. HME (trivial) is for symmetric configurations where the toroidal field remains zero in the core according to hydromagnetic equilibrium. }
\label{tab:example}
\begin{tabular}{lccccc}
\hline
Model & Crust Evolution & Core Poloidal Evolution & Core Toroidal Evolution & $|\vec{B}|_\text{max}\text{ (initial)}$ & Section \\
\hline
A & Hall + Ohmic & Fixed & HME (Relaxation) & $2.2\times10^{14}$ G & Section \ref{hydromagnetic}\\
B & Hall + Ohmic & Jones drift only & HME (Trivial) & $2.2\times10^{14}$ G & Section \ref{Jones_strong_B}\\
C & Ohmic & Jones drift only & HME (Trivial) & $2\times10^{12}$ G & Section \ref{Jones_weak_B}\\
D1 & Hall + Ohmic & Jones drift + Vortex-transport & HME (Trivial) & $6.9\times10^{12}$ G & Section \ref{vortex_expulsion}\\
D2 & Hall + Ohmic & Jones drift + Vortex-transport  & HME (Trivial) & $1.4\times10^{13}$ G & Section \ref{vortex_expulsion}\\
D3 & Hall + Ohmic & Jones drift + Vortex-transport  & HME (Trivial) & $2.7\times10^{13}$ G & Section \ref{vortex_expulsion}\\
E & Hall + Ohmic + Elastic & Fixed & HME (Trivial) & $2\times10^{14}$ G & Section \ref{magneto-elastic}\\
\hline
\end{tabular}
\label{results_table}
\end{table*}
In conventional neutron star models, neutrons in the core form cooper pairs, and exist in a $^3P_2$ superfluid state \citep{baym_neutron_1975}. The vorticity of the bulk fluid must be zero, and circulation is quantized on microscopic scales with the formation of superfluid vortices. The vortices each possess the quantum of circulation $\kappa=h/2m_n$, where $h$ is the Planck constant and $2m_n$ is the effective mass of a cooper pair. The mean number density of vortices is
\begin{equation}
n_v = \frac{2\Omega_n}{\kappa},
\end{equation}
where $\Omega_n$ is the superfluid rotational frequency. The neutron vortices are not necessarily straight, though the absence of a firm detection of free precession seems to indicate that the vortex configuration is not radically different from a linear array. 

Stellar spin-down must be accompanied by motion of these vortices outward, in order to conserve angular momentum \citep{ruderman_rotating_1974}. Likewise spinning up the star must be accompanied by motion of the neutron vortices inward. The neutron vortices move in the radial direction with velocity
\begin{equation}
\vec{v}_\perp = -\frac{r_\perp\dot{\Omega}_n}{2\Omega_n}\vec{\hat{e}}_{r_\perp}.
\label{vperp}
\end{equation}
As a consequence of the neutron superfluid coupling to the proton superfluid, protons also circulate around the neutron vortices, which produces a magnetization localized within the penetration depth $\lambda\lesssim10^{-11}$ cm of the neutron vortex core [see eg. \cite{jones_neutron_1991}]. Outward moving neutron vortices interact strongly with flux tubes, and thus the spin-down of neutron stars can result in the transport of magnetic flux tubes \citep{srinivasan_novel_1990}. According to \cite{ruderman_neutron_1998}, force builds up on the flux tubes, which are either carried along with the neutron vortices, or are cut through by them. \cite{jones_type_2006} states that in order for straight flux tubes to be pushed along by neutron vortices (with velocity $\vec{v}_\perp$), the maximum velocity of the flux tubes (due to viscous drag) must be greater than or equal to the velocity of the neutron vortices, $|\vec{v}_{F}|\geq|\vec{v}_\perp|$. If this inequality is not satisfied, the neutron vortices will cut through the flux tubes. \cite{jones_type_2006} gives the maximum velocity of flux tubes being pushed on by neutron vortices as 
\begin{equation}
\vec{v}_F = \alpha n_v \tilde{f}_v \vec{\hat{e}}_{r_\perp},
\end{equation}
where $\alpha$ is the drag coefficient defined in Equation~\eqref{jones_v}, and $\tilde{f}_v$ is the maximum force per unit length a neutron vortex can exert on a flux tube without cutting through. We estimate $\tilde{f}_v$ by using the vortex-flux tube interaction energy of \cite{gugercinoglu_microscopic_2016}\footnote{These authors have corrected a typo in the interaction energy of \cite{ruderman_neutron_1998} which erroneously leads to a factor of $\pi^2$ larger pinning energies.}. Following \cite{ruderman_neutron_1998} we assume that as a vortex moves along and collects flux tubes, the separation between flux tubes approaches $2\lambda$. This gives
\begin{equation}
\tilde{f}_v = \frac{\phi_0\phi_0^*}{8\pi^2\lambda^3}\text{ln}\left(\frac{\lambda}{\xi_p}\right)\approx 3.8\times 10^{17} \text{ dyne/cm},
\end{equation}
with $\phi_0^*$ the flux quantum of a neutron vortex (we approximate $\phi_0\approx\phi_0^*$), and $\xi_p$ the proton coherence length.

We extend the treatment of \cite{jones_type_2006} and \cite{ruderman_neutron_1998} to the case of curved flux tubes, by including the self-tension force $\vec{f}_B$, and also discuss the transport of flux tubes in the cut-through regime. We say that cut-through occurs when
\begin{equation}
\left| -\frac{\vec{v_\perp}}{\alpha}\cdot\hat{\vec{e}}_{r_\perp} + \vec{f_B}\cdot\hat{\vec{e}}_{r_\perp} \right| \geq n_v\tilde{f}_v,
\label{cut_through}
\end{equation}
with $-\vec{v_\perp}\cdot\hat{\vec{e}}_{r_\perp}/\alpha$ the drag force the flux tubes exert on the neutron vortices (assuming a stationary background fluid), $\vec{f_B}\cdot\hat{\vec{e}}_{r_\perp}$ the tension force flux tubes exert on the vertical vortices, and $n_v\tilde{f}_v$ the maximum force per unit volume the neutron vortices can exert on the flux tubes. We call satisfaction of the above inequality ``cut-through", and dissatisfaction of the inequality ``vortex-transport". \\

In the transport regime, the flux tubes are carried along with velocity $\vec{v}_\perp$. But the flux tubes also have the freedom to slide along the neutron vortices, with the projected Jones drift velocity $(\vec{v}_\text{J}\cdot\hat{\vec{e}}_n)\hat{\vec{e}}_n$. Here $\hat{\vec{e}}_n$ is a unit vector which points along the local direction of a vortex.\\

In the cut-through regime, we assume that the flux tubes are still carried along by the neutron vortices, but only at the terminal velocity $\vec{v}_F$. Since the vortices cannot prevent the motion of flux tubes in the cut-through regime, the flux tubes can also drift in accordance with their own self tension (the Jones drift velocity $\vec{v}_\text{J}$). To summarize the flux tubes are advected with the velocity field
\begin{equation}
\vec{v_\text{sd}} = 
\begin{cases}
    \vec{v}_\perp + (\vec{v}_\text{J}\cdot\hat{\vec{e}}_n)\hat{\vec{e}}_n,& \text{(vortex-transport)}\\
    \vec{v}_F + \vec{v}_\text{J},  & \text{(cut-through)}
\end{cases}
\label{v_sd}
\end{equation}
while the neutron vortices move with velocity $\vec{v}_\perp$. We note here that the velocity field given by Equation~\eqref{v_sd} is in fact a piece-wise continuous function. This can be understood by noting that if the LHS of Equation~\eqref{cut_through} is slightly greater than the RHS, then the velocity there is set to $\vec{v}_F+\vec{v}_\text{J}$, so that a discontinuity never develops. See \cite{ruderman_neutron_1998} for a similar model of vortex transport in 1D for the case of straight flux tubes.

We assume that the neutron vortex array is not significantly deformed by the flux tubes. This is true for high spin frequencies, when the neutron magnus force is larger than the critical cut-through force
\begin{equation}
\Omega >\frac{\tilde{f}_v}{ r_\perp \rho_n \kappa}.
\end{equation}

The evolution of the magnetic field due to motion of neutron vortices is given by
\begin{equation}
\frac{\partial \vec{B}_\text{p}}{\partial t} = \nabla\times(\vec{v}_\text{sd}\times\vec{B}_\text{p}).
\end{equation}
In terms of the scalar function $\Psi$, the evolution is
\begin{equation}
\frac{\partial\Psi}{\partial t} + \vec{v}_\text{sd}\cdot\nabla\Psi=0,
\label{vortex_eqn}
\end{equation}
while the toroidal field evolves according to hydromagnetic equilibrium. 

\section{Results}\label{results}
The equations describing the magnetic field evolution in the star, i.e. the evolution equations for functions $\Psi(r,\theta)$ and $I(r,\theta)$, are discretized on a grid and solved numerically as described in Appendix \ref{AppendixB}.

\subsection{Hydromagnetic core} \label{hydromagnetic}

\begin{figure*}
\centering
\subfloat{\includegraphics[width=1.\textwidth]{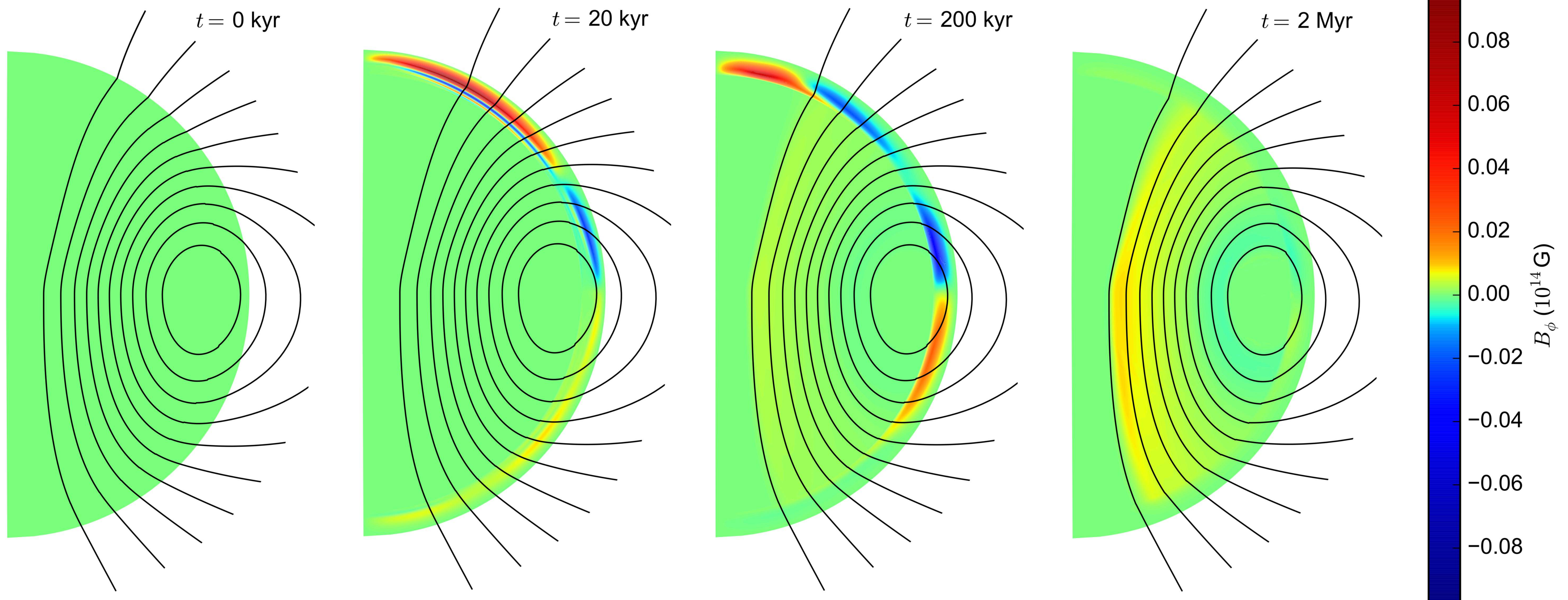}}
\caption{Snapshots of the magnetic field evolution for Model A (Table \ref{results_table}), shown at $t=0$, $20$ kyr, $200$ kyr, and $2$ Myr. The black curves are 10 contour lines of the poloidal flux function $\Psi$ (i.e. the polodial magnetic field lines), equally spaced between $\Psi=0$ and the maximum value $\Psi_\text{max}$, at $t=0$. The toroidal field is represented by the colour scale, which varies logarithmically, with a linear region around zero.}
\label{HME_core}
\end{figure*}

\begin{figure}
\centering
\subfloat{\includegraphics[width=.5\textwidth]{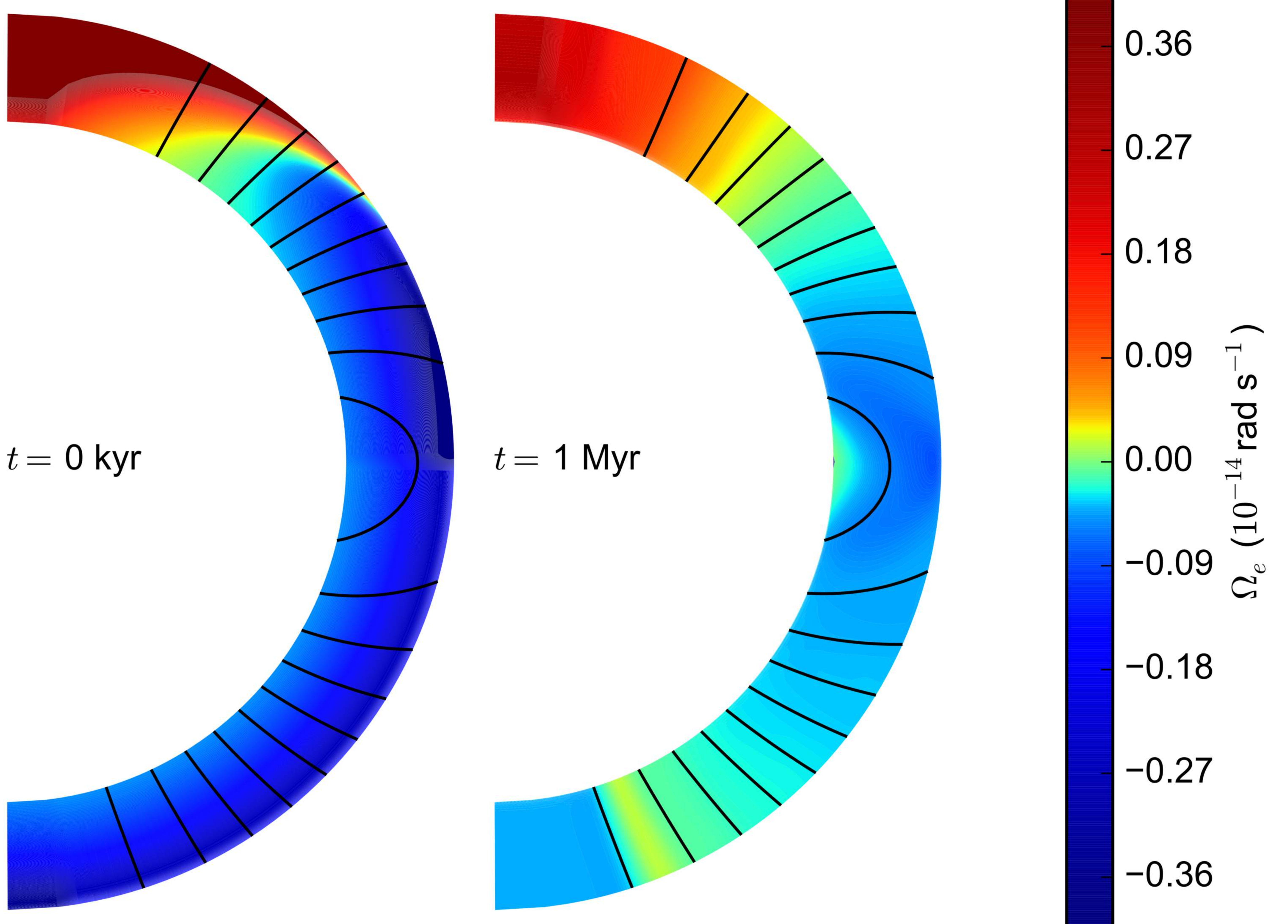}}
\caption{Snapshots of the evolution for Model A (Table~\ref{results_table}), shown at $t=0$ kyr, and $1$ Myr. The plotting scheme is the same as Figure \ref{HME_core} but here colour shows $\Omega_\text{e}$ (the angular velocity of the electron fluid), the result of Hall drift in the crust. The thickness of the crust has been magnified by a factor of 2.5.}
\label{omega_hall}
\end{figure}
First we consider Model A in which the effects of stellar spin-down, and Jones flux tube drift, and ambipolar diffusion are neglected (ie. the poloidal field is static in the core). The purpose of this section is to demonstrate clearly the hydromagnetic equilibrium described in Section~\ref{hydromagnetic_theory}. The drift of poloidal field lines in the core will be studied in Sections~\ref{JonesDrift} and \ref{vortex_expulsion}.  

We note first that a significant twist angle can be associated with toroidal fields which have comparable strength to the poloidal field. For a field line in the core of length $L$, we estimate the twist angle to be of order
\begin{equation}
\begin{split}
\zeta \sim \frac{L B_\phi}{r_\perp B_\lambda} \sim \left(\frac{L}{10^{6}\text{ cm}}\right)\left(\frac{10^{6}\text{ cm}}{r_\perp}\right)\frac{B_\phi}{B_\lambda}\sim 1\text{ rad},
\end{split}
\end{equation}
when $B_\phi\approx B_\lambda$. Changing the twist angle of a field line in the core requires differential rotation of its two ends where it is attached to the crust. Equatorial symmetry of the magnetic field implies no differential rotation -- the two ends must move with the same speed. The same fact is seen formally from the equations. For the case of equatorial symmetry, the net flux of twist into the core vanishes for each field line, as 
\begin{equation}
J(\Psi,\lambda_2,t)-J(\Psi,\lambda_1,t)=0,
\end{equation}
(see Section \ref{hydromagnetic_theory}). If the initial field has plane reflection symmetry about the equator, it will maintain this symmetry throughout the evolution. Then, by Equation \eqref{coreev}, we see that
\begin{equation}
\partial_t  I_\text{core}(\Psi,t)=0 \quad \text{(if symmetric)},
\end{equation} 
for all time. In reality it is likely that young neutron stars will have some toroidal field in order to stabilize the poloidal field. However in some of our simulations we consider a number of initial fields which display equatorial reflection symmetry, and have purely poloidal fields in the core. In this special case the evolution of the core toroidal field is trivial - it remains zero according to hydromagnetic equilibrium. 

Poloidal fields which violate equatorial plane reflection symmetry give a non-zero net flux of twist into the core, 
\begin{equation}
J(\Psi,\lambda_2,t) - J(\Psi,\lambda_1,t)\neq0,
\end{equation}
and we may see evolution of the toroidal field in the core. With this in mind we choose the initial poloidal field in Figure~\ref{HME_core} to violate equatorial reflection symmetry. We enforce hydromagnetic equilibrium in Model A with the relaxation method outlined in Section~\ref{hydromagnetic_theory}. 

\begin{figure*}
\centering
\subfloat{\includegraphics[width=1.\textwidth]{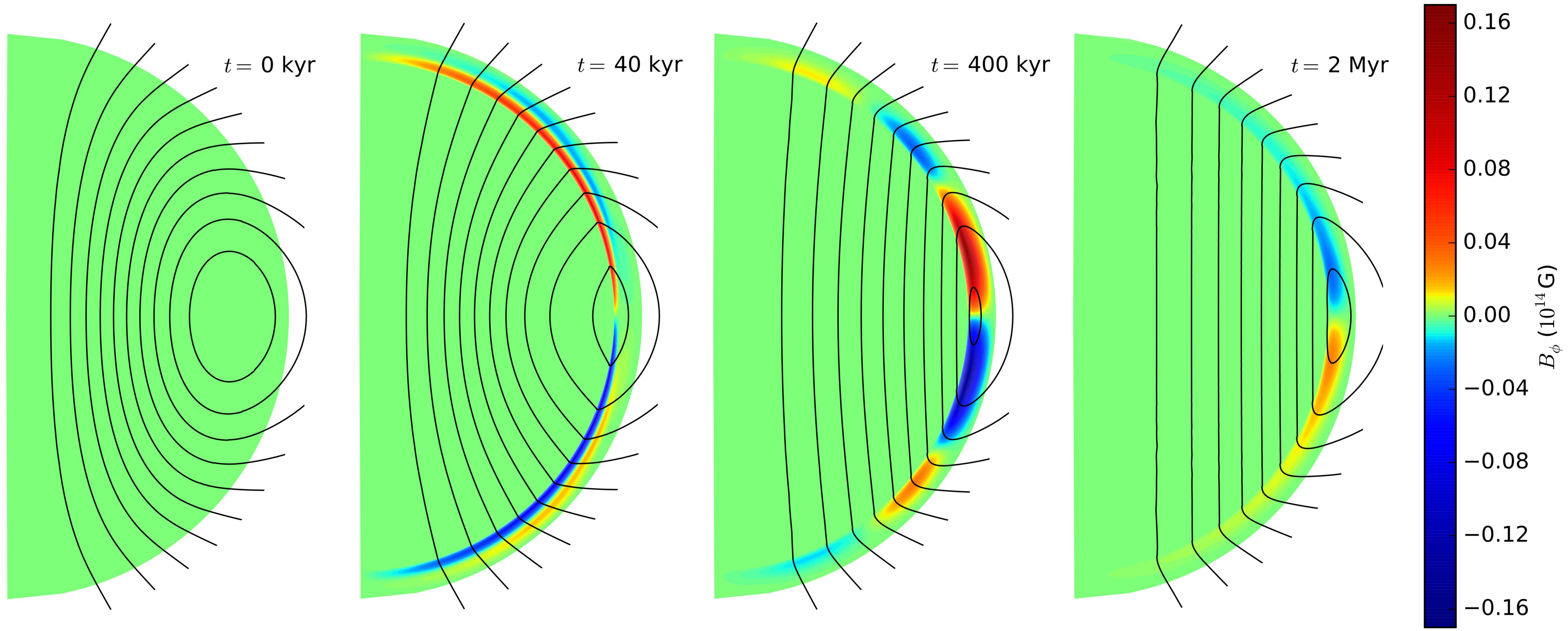}}
\caption{Snapshots of the magnetic field evolution for Model B (Table~\ref{results_table}), shown at $t=0$~kyr, $40$~kyr, $400$~kyr, and $2$ Myr. The plotting scheme is the same as Figure \ref{HME_core}.}
\label{jones_magnetar}
\end{figure*}

The simulation shown in Figure~\ref{HME_core} starts with a current sheet on the surface of the star in the northern hemisphere (see Figure~\ref{omega_hall}). The current sheet shears poloidal field lines near the surface, and generates toroidal field with positive polarity in the northern hemisphere, and negative in the south. Hall drift in the crust slowly winds the core magnetic field in the azimuthal direction. At $t=200$ kyr there is a weak toroidal field in the core, and several patches of toroidal field in the crust with alternating polarity. After $t\sim 600$ kyr the toroidal field reaches a steady state, with two patches near the equator in opposite hemispheres, which are damped by Ohmic diffusion from this point on. By this time the poloidal field has settled into the Hall attractor state, corresponding to constant electron angular drift velocity along poloidal field lines in the crust ($\Omega_\text{e} = \Omega_\text{e}(\Psi)$). Hall drift in the crust continues to wind the core field in the azimuthal direction, though more slowly as Ohmic diffusion dissipates the crustal currents. At $t=2$ Myr the core supports a toroidal field of strength $B\sim 10^{12}$ G, and similar in the crust. The twist angle of a typical field line in the core is of order $10^{-2}\text{ rad}$, so Hall drift has only weakly twisted the core magnetic field in this model. It is possible that for different initial conditions, stronger azimuthal currents could inject a larger twist into the core. This model has confirmed the Hall attractor of \cite{gourgouliatos_hall_2014} for core penetrating B-fields, with the correct hydromagnetic equilibrium enforced.

\subsection{Jones flux tube drift} \label{JonesDrift}

\subsubsection{Dissipative straightening of flux tubes}
In a more realistic model the poloidal field lines in the core are not fixed, and are expected to drift as outlined in Section \ref{Jones_theory}. Jones drift allows flux tubes to straighten by slipping with some viscous dissipation through the core electron fluid. As the flux tubes straighten, free energy stored in the curvature of the flux tubes is dissipated. The characteristic timescale of this straightening is 
\begin{equation}
t_\text{diss}\sim\frac{s}{v_\text{J}}
\end{equation}
with $s$ the deviation from straight flux tubes. We approximate a curved flux tube as a circular arc, with $s\approx r_c^2/2 R_c$, and radius of curvature $R_c$. The Jones drift velocity is approximated as
\begin{equation}
\begin{split}
v_\text{J} &\sim \frac{\tilde{\sigma}}{4\pi n_e^2 e^2}\frac{BH_{\text{c1}}}{R_c} \\&=2.8\times10^{-8}\left(\frac{3.5\times 10^{37} \text{ cm}^{-3}}{n_e} \right)^2\left(\frac{\tilde{\sigma}}{10^{29}\text{ s}^{-1}} \right) \text{cm s}^{-1}.
\end{split}
\label{velocity}
\end{equation}
The timescale for flux tubes to straighten is then
\begin{equation}
\begin{split}
t_\text{diss} & \sim \frac{2\pi n_e^2 e^2}{\tilde{\sigma}}\frac{r_c^2}{BH_\text{c1}} \\& = 450 \left(\frac{n_e}{3.5\times 10^{37} \text{ cm}^{-3}} \right)^2\left(\frac{10^{29}\text{ s}^{-1}}{\tilde{\sigma}} \right) \text{ kyr},
\end{split}
\end{equation}
where we have taken the estimate of \cite{jones_type_2006} $\tilde{\sigma} = 10^{29} B_{12}^{-1}\text{s}^{-1}$. Note that the timescale is independent of the field strength $B$. The timescale for straightening can also be significantly shorter than the above estimate, depending on $\tilde{\sigma}$, which can be larger for cores with high muon densities \citep{jones_type_2006}. Our advection velocity in Equation~\eqref{velocity}, illustrates the discrepency in the timescales of \cite{elfritz_simulated_2016} and \cite{jones_type_2006}. The advection velocity of \cite{elfritz_simulated_2016}, is typically $\sim 10^{-11}\text{cm s}^{-1}$ [\cite{elfritz_simulated_2016}, Figures 5, 8, and 11 therein].

The Jones drift velocity acts perpendicular to poloidal field lines, in order to minimize the curvature. Thus, Jones drift becomes inactive when the flux tubes are straightened. However, when the field is straightened in the core, a sharp cusp forms on the crust-core interface, supported by strong toroidal currents at the base of the crust. This cusp will therefore be site to rapid Ohmic diffusion, which smooths the cusp, generating curvature in the field lines inside the core, and reactivating the Jones effect, which proceeds to straighten them again. So we see that the coupled crust-core system continuously evolves under the combined effects of Jones flux tube drift in the core, and Ohmic diffusion in the crust. 

In this section we explore the evolution in two scenarios. Firstly, we consider the drift of flux tubes in a strongly magnetized neutron star, with Hall drift and Ohmic diffusion active in the crust. Secondly, we study the long timescale evolution of a moderately magnetized neutron star, and determine the decay timescale.

\subsubsection{Flux tube drift and Hall drift (strong B)}\label{Jones_strong_B}
For strong magnetic fields, Hall drift can interfere with the flux tube drift in the core (when $t_\text{Hall}$ is comparable to $t_\text{diss}$). Figure~\ref{jones_magnetar} shows the evolution of a highly magnetized neutron star (Model B) with an initially poloidal field, evolving by Jones' flux tube drift in the core, coupled to a crust evolving with Hall drift and Ohmic diffusion. For this simulation we use $\tilde{\sigma} = 10^{29}B_{12}^{-1}\text{s}^{-1}$. The initial field has maximum strength $B\approx 3\times 10^{14}$~G in the core. The initial field displays equatorial symmetry, and thus Hall drift will not inject any magnetic twist into the core, meaning that the toroidal field remains zero there. 

There are two main stages to the evolution of the core magnetic field in Model B. The first stage lasts for $t_\text{diss}$, and involves a rapid straightening of the flux tubes in order to relieve magnetic stresses. During this stage the core field dissipates its initial free energy, on viscous slippage through the fluid. The straightening of flux tubes in the core is associated with the formation of a sharp cusp in the poloidal field at the crust-core interface, supported by a toroidal current sheet at the base of the crust. This current sheet generates toroidal field deep in the crust through the Hall effect. The current sheet is also site to enhanced Ohmic dissipation. The regions of toroidal field in the crust are advected toward the equator, and much weaker higher order multipole structure forms in the toroidal field which is efficiently damped. The toroidal field is sufficiently weak, that it does not cause any large scale rearrangement of the poloidal field. Eventually activity due to Hall drift declines, as the Hall attractor drives the poloidal field in the crust to a state of rigid rotation of the electron fluid.

The second stage begins at $t\sim 1$~Myr. During this stage, Ohmic diffusion controls the evolution, which becomes a self-similar decay of the global magnetic field. Jones drift allows the core magnetic field to adjust on a timescale which is faster than Ohmic diffusion at the base of the crust, so that they effectively remain straight for the remainder of the evolution. Flux tubes in the core gradually drift outward, consistent with the rate of Ohmic diffusion at the base of crust. Analytic estimates describing this ``Ohmic drift" of straight field lines will be given in Section \ref{Jones_weak_B}, where we consider this drift in isolation, in the absence of any Hall drift. Flux in the core converges toward the null point in the field, which is located at the equator, on the crust-core interface, for this particular configuration. At the null point, the field lines close and annihilate. The evolution of the field into this state of self similar decay is not unique to these initial conditions, and we observe the same final state for a number of approximately dipole initial magnetic fields. We note that the Jones drift timescale does not scale with field strength, and likewise with Ohmic diffusion. This implies that timescales relating to the evolution of poloidal fields in Model B could be applied to initial fields with a variety of strengths.

\subsubsection{Flux tube drift and no Hall drift (moderate B)}\label{Jones_weak_B}
\begin{figure*}
\centering

\subfloat{\includegraphics[width=1.\textwidth]{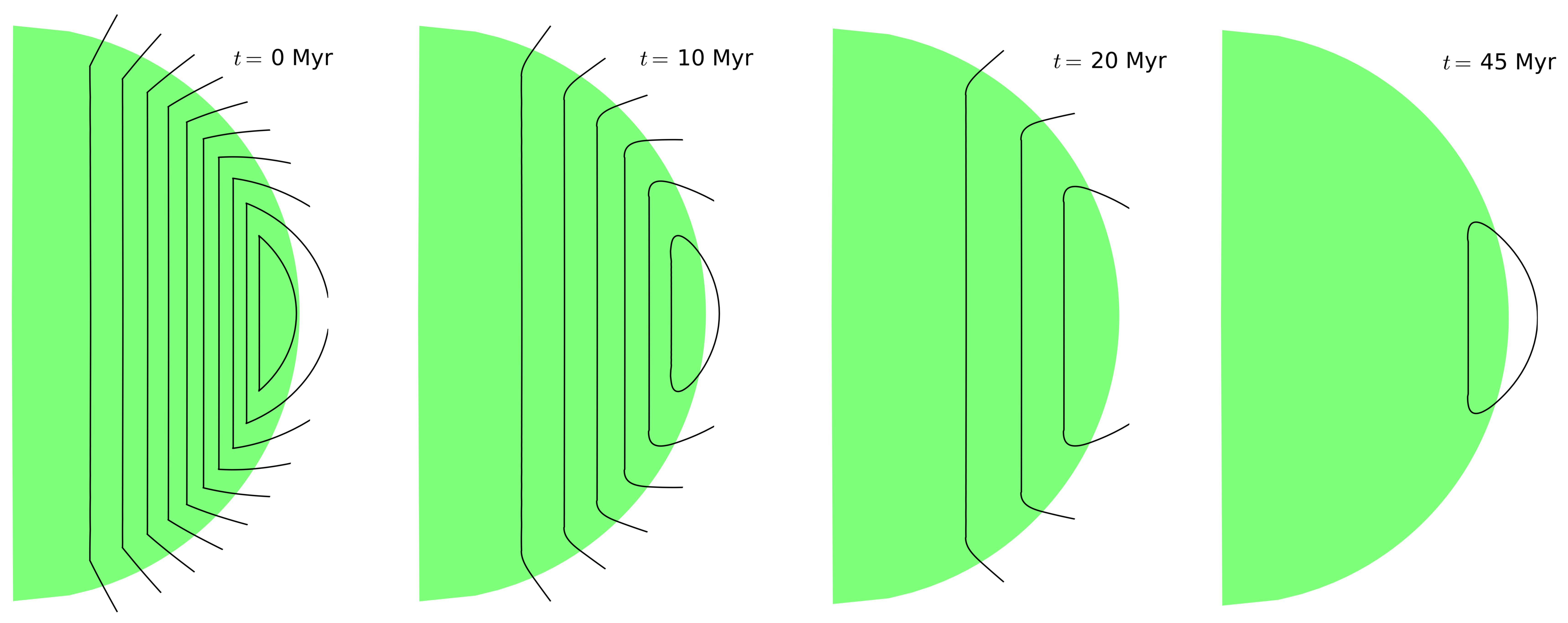}}
\caption{Snapshots of the magnetic field evolution for Model C (Table~\ref{results_table}), shown at $t=0$~Myr, $10$~Myr, $20$~Myr, and $45$~Myr. The plotting scheme is the same as Figure \ref{HME_core}. The toroidal field is everywhere zero.}
\label{jones_pulsar}
\end{figure*} 
The second scenario of interest is the evolution of pulsar strength magnetic fields due to Jones flux tube drift in the core, in the case where crustal Hall drift is not important. This regime occurs for typical pulsar fields of $B\sim 10^{12}$~G or lower. With this in mind, in Model C we consider the crust evolving under Ohmic diffusion only so that long timescale simulations are less computationally expensive. We also avoid evolving the flux tubes in the core directly [Equation \eqref{jones_eqn}], and instead enforce the boundary condition at the base of the crust that field lines remain vertical in the core ($B_\theta = B_r\text{tan}\theta$ at $r=r_c$) as a result of Jones drift. This is a good approximation because the flux tubes can always straighten faster than Ohmic diffusion at the base of the crust, as seen in Figure~\ref{jones_magnetar}. 

\begin{figure}
\centering
\subfloat{\includegraphics[width=.4\textwidth]{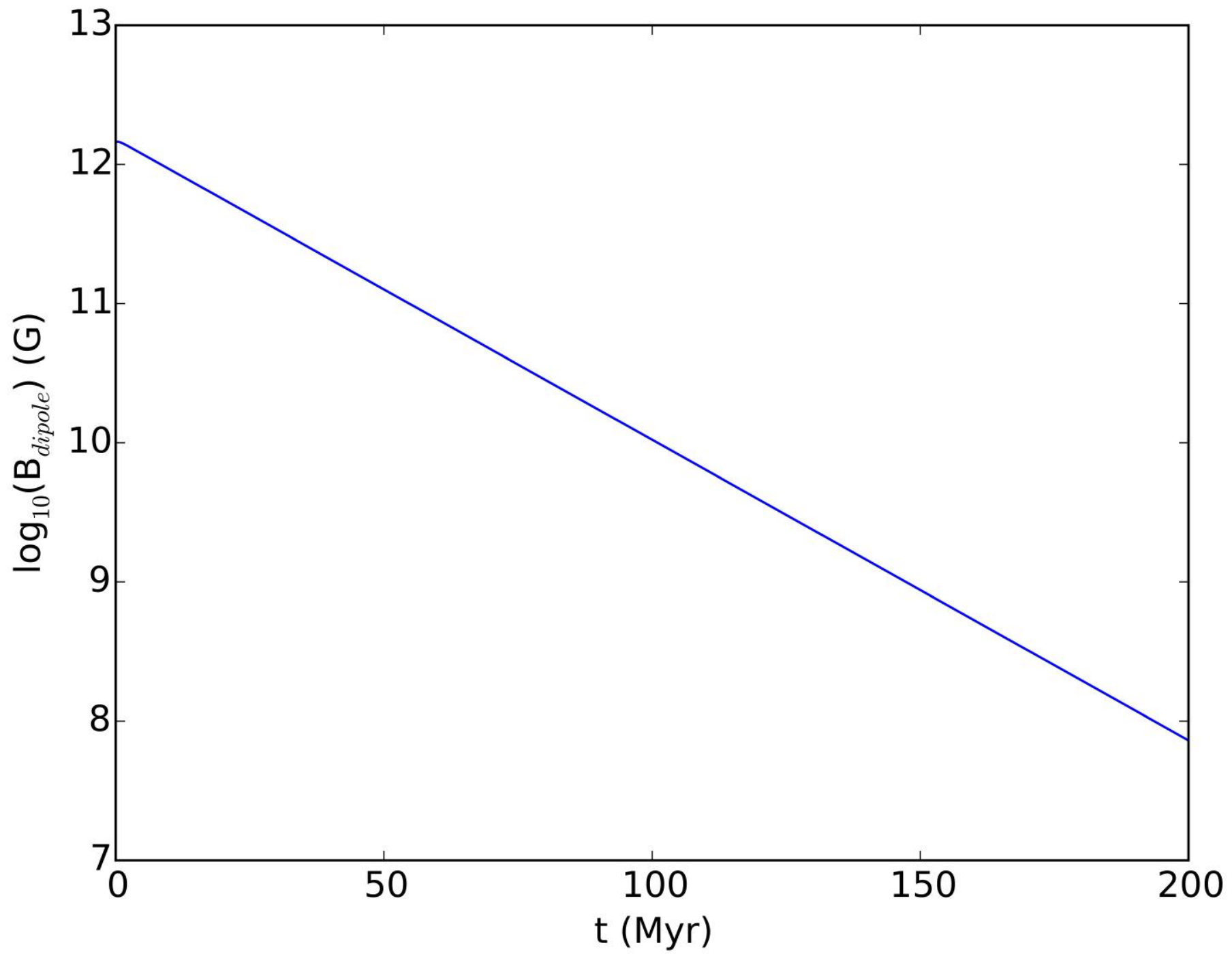}}
\caption{Decay of the dipole field for the magnetic field evolution shown in Figure~\ref{jones_pulsar}.}
\label{dipole_plot}
\end{figure}

Figure~\ref{jones_pulsar} shows the long timescale evolution of Model~C, with Jones flux tube drift in the core, coupled to a crust with Ohmic diffusion. The field in the core is pure $B_z$ (In particular $\Psi\propto r_\perp^2$), and the initial crustal field is a dipole potential field matched on to the core. The field has typical strength $B\approx10^{12}$ G. In the first $\sim 1\text{ Myr}$, diffusion at the base of the crust smooths the kink in the poloidal field, and the crustal field relaxes into an Ohmic eigenmode. From this point on the evolution of the global field can be likened to self similar decay. Tension in the magnetosphere ensures that poloidal field lines in the crust converge toward the null point at the equator. The field lines in the core are pulled along at the rate set by Ohmic diffusion, also toward the null point at the base of the crust, where they close and annihilate. For the remainder of the evolution the structure of the magnetic field remains unchanged, as it gradually grows weaker. The evolution of the dipole field strength is plotted in Figure~\ref{dipole_plot}. After $\sim150$ Myr, the dipole field strength has decreased from $B\approx10^{12}$~G to $B\approx10^{9}$ G.

\begin{figure}
\centering
\subfloat{\includegraphics[width=.4\textwidth]{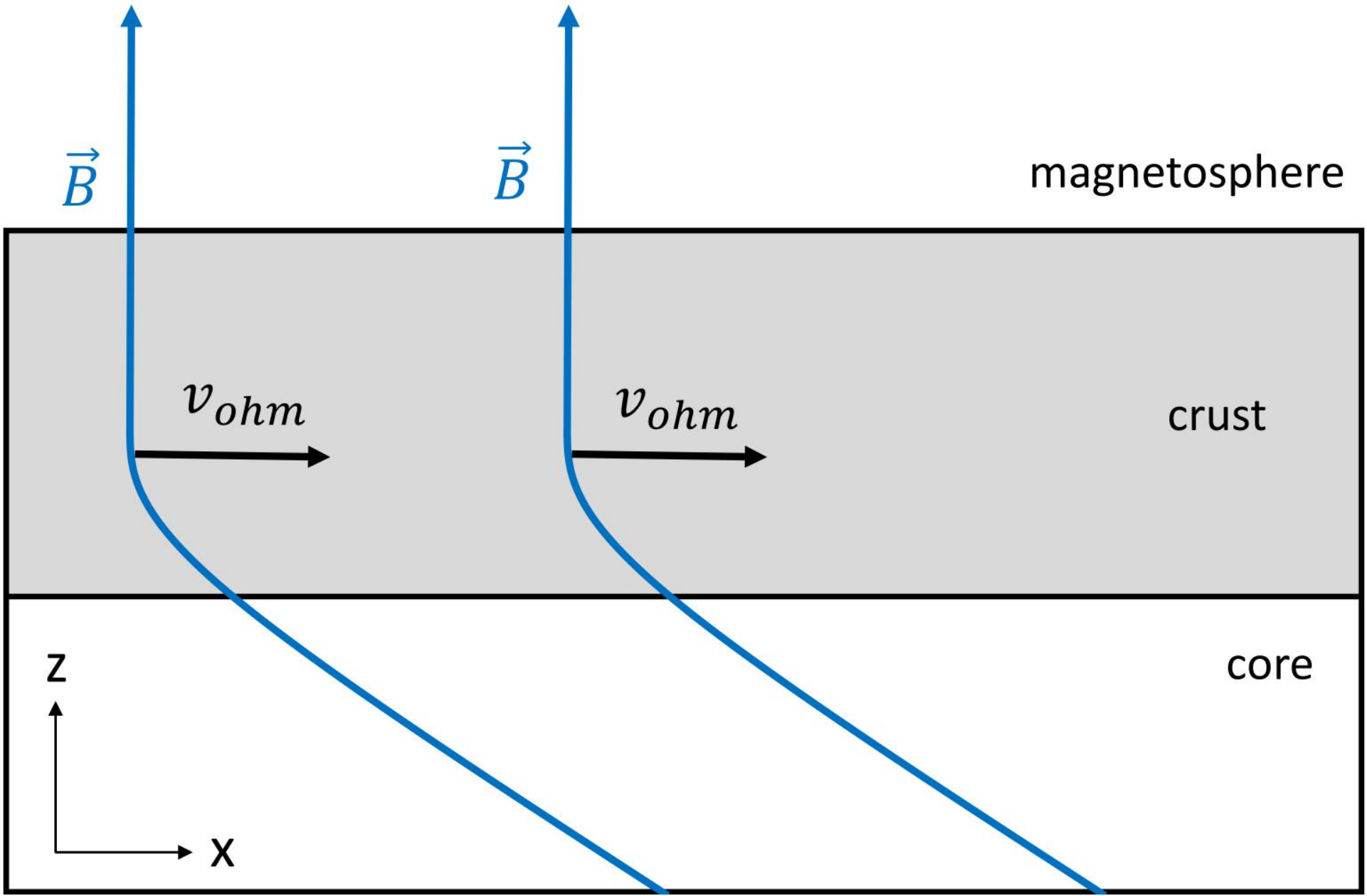}}
\caption{Plane parallel slab (grey), with length $l$. Field lines (blue) move to the right consistent with the rate set by Ohmic diffusion.}
\label{plane_crust}
\end{figure}

The timescale for magnetic flux to diffuse through the crust in the above scenario is very different from the Ohmic timescale of \cite{goldreich_magnetic_1992}. As an approximation, consider the cartesian configuration shown in Figure~\ref{plane_crust}, and assume a constant diffusivity $\eta$ throughout the crust. Curvature of the field lines at the base of the crust generates a current sheet of thickness $h$, given by Ampere's law as
\begin{equation}
j \sim -\frac{c}{4\pi}\frac{B_x}{h},
\end{equation}
using Ohm's law this gives the electric field
\begin{equation}
E \sim -\frac{c}{4\pi\sigma}\frac{B_x}{h}.
\end{equation}
As in Section~\ref{hydromagnetic_theory}, we use this electric field to estimate the velocity of magnetic field lines due to Ohmic diffusion through $\vec{E} = -\vec{v} \times \vec{B}/c$. The resulting velocity of field lines in the x-direction is
\begin{equation}
v_\text{ohm} \sim \frac{\eta}{h}\frac{B_x}{B_z},
\end{equation}
with $B_z$ and $B_x$ the vertical and horizontal components of the field. For this configuration a quasi-steady drift is established with the current sheet occupying the region of the crust with highest conductivity, i.e. the deep crust. Its thickness $h$ is a few hundred meters. The quasi-steady drift is established on the timescale $h^2/\eta$, and the drift is associated with the transport of magnetic field lines with characteristic time
\begin{equation}
\begin{split}
\tilde{t}_\text{ohm} &\sim \frac{hl}{\eta}\frac{B_z}{B_x} \\& = 150\left(\frac{h}{3\times 10^{4}\text{ cm}}\right)\left(\frac{l}{\pi r_*}\right)\left(\frac{\sigma}{3.6\times 10^{24}\text{ s}^{-1}}\right)\text{Myr},
\end{split}
\label{modified_ohmic_time}
\end{equation}
where we have assumed $B_z\approx B_x$, and the current sheet thickness $h\approx 3\times 10^4\text{ cm}$ corresponding to the highly conducting region of the deep crust. We have used $\sigma = 3.6\times10^{24}$ s$^{-1}$, corresponding to phonon scattering at $T\approx2\times10^8$ K \citep{gourgouliatos_hall_2014}. This timescale can be greater by an order of magnitude compared to the Ohmic timescale of \cite{goldreich_magnetic_1992}, depending on the thickness of the crust and the geometric factor $B_z/B_x$. The timescale given by Equation~\eqref{modified_ohmic_time} corresponds to the time taken for the dipole field strength to decay by approximately 3 orders of magnitude for the numerical simulation seen in Figure~\ref{jones_pulsar}.

\begin{figure*}
\centering

\subfloat{\includegraphics[width=.8\textwidth]{figure_7}}

\caption{Snapshots of the magnetic field evolution for Model D1 (Table~\ref{results_table}), shown at $t=0$ kyr, $5$ kyr, $100$ kyr, $1$ Myr, $2$ Myr, and $4$ Myr. The plotting scheme is the same as Figure \ref{HME_core}. This simulation begins $300$ yr after the neutron star birth, with surface field strength $\sim5\times10^{12}$~G, and spin period $10.9$~ms.}
\label{n_vortex1}
\end{figure*}

\begin{figure*}
\centering

\subfloat{\includegraphics[width=.8\textwidth]{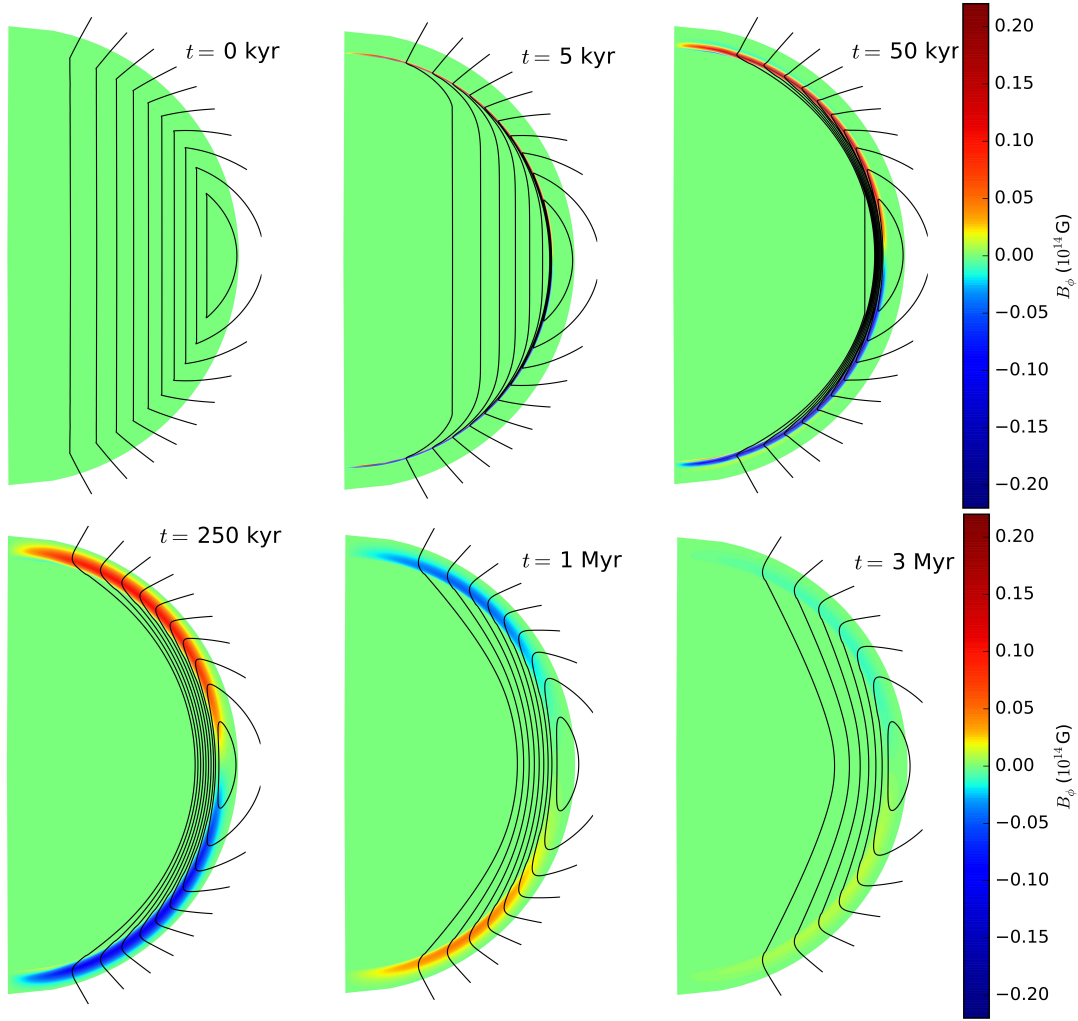}}

\caption{Snapshots of the magnetic field evolution for Model D2 (Table~\ref{results_table}), shown at $t=0$ kyr, $5$ kyr, $50$ kyr, $250$ kyr, $1$ Myr, and $3$ Myr. The plotting scheme is the same as Figure \ref{HME_core}. This simulation begins $300$ yr after the neutron star birth, with surface field strength $\sim 10^{13}$~G, and spin period $21.8$~ms.}
\label{n_vortex2}
\end{figure*}

\begin{figure*}
\centering

\subfloat{\includegraphics[width=.8\textwidth]{figure_9}}

\caption{Snapshots of the magnetic field evolution for Model D3 (Table~\ref{results_table}), shown at $t=0$ kyr, $5$ kyr, $50$ kyr, $150$ kyr, $1$ Myr, and $3$ Myr. The plotting scheme is the same as Figure \ref{HME_core}. This simulation begins $300$ yr after the neutron star birth, with surface field strength $\sim 2\times10^{13}$~G, and spin period $43.5$~ms.}
\label{n_vortex3}
\end{figure*}

\subsection{Rapidly rotating newborn neutron stars}\label{vortex_expulsion}
The origin of strong magnetic fields in neutron stars is not well understood. Some models suggest a fossil field, left behind by the progenitor \citep{ferrario_modelling_2006}, while others invoke dynamo mechanisms requiring the neutron star to be born with millisecond spin periods \citep{thompson_neutron_1993}. Here we assume that a highly magnetized neutron star can be born with a $1$ millisecond spin period, and consider the implications for the evolution of the magnetic field. In particular, we are interested in the interaction of superfluid neutron vortices with flux tubes, as a means for the rotational energy of the star to rearrange the core magnetic field. Importantly the cooling curves of the Cas A remnant suggest that the transition to superfluidity in the core takes place after $t\sim300$ years [\cite{shternin_cooling_2011}, \cite{page_rapid_2011}]. As was pointed out by \cite{thompson_global_2017}, by this time a typical magnetar ($B\sim10^{15}$ G) will have a spin period $>1$ s. It is simple to show that the ratio of rotational to magnetic energy is very small, and it is difficult to envisage how the core field could be significantly rearranged by a transfer of rotational energy. However, for neutron stars born with weaker magnetic fields, corresponding to a so-called weak-field magnetar (or high-B pulsar), this is not the case. Specifically, if a neutron star is born with initial spin period $1$ ms, and $B\sim10^{13}$ G, then after $300$ years, the spin period will be $21.8$ ms. Then the ratio of rotational to magnetic energy is large, and will remain so for an extended period after the transition to neutron superfluidity. 

In Models D1, D2 and D3 we model neutron stars, with a range of initial magnetic field strengths, assuming a birth spin period $1$ ms. Our simulations begin $300$ yr after the neutron star birth, corresponding to the time of phase transition to neutron superfluidity. We model the spin evolution of the star self consistently according to $\dot{\Omega} = -\beta\Omega^3$, where $\beta = 2a_1^2/3c^3I$. Here $a_1$ is the dipole moment of the surface magnetic field, and we take $I=10^{45}$ g cm$^2$ as a typical moment of inertia. The spin evolution of the star determines the velocity of the neutron vortices, according to
\begin{equation}
\vec{v}_\perp = -\frac{r_\perp\dot{\Omega}_n}{2\Omega_n}\vec{\hat{e}}_{r_\perp}.
\label{vperp2}
\end{equation}

In Model D1 we use an initial field of strength $\sim5\times10^{12}$ G at the surface, and $6.9\times10^{12}$ G in the core. We set the initial (here initial refers to the beginning of the simulation--$300$ yr after the neutron star birth) spin period accordingly to $10.9$ ms. We show the results of this simulation in Figure~\ref{n_vortex1}. In the first $100$ kyr the flux is rapidly pushed out of the core at the velocity $\vec{v}_\perp$. In the outer core the flux tubes are severely deformed, and the tension force $\vec{f}_B$ becomes large enough to cause cut-through in a thin layer beneath the crust. The sharp curvature of poloidal field lines at the base of the crust is site to a strong toroidal current sheet, which generates a quadrupolar toroidal field through Hall drift, and rapid Ohmic dissipation of the poloidal field. The expulsion of flux tubes can result in an order of magnitude increase in the poloidal field strength in the outer core.

From this point on, the flux tubes slide vertically along the neutron vortices toward the equator (away from the crust-core interface) with the projected Jones velocity $(\vec{v}_\text{J}\cdot\hat{\vec{e}}_n)\hat{\vec{e}}_n$. The flux tubes form a ``$>$" shape, with the cusp located along the equator in the outer core. The cusp becomes sharper, until the tension force $\vec{f}_B$ is large enough to cut through the vortices, and begins to minimize it at $\sim 880$ kyr. After $\sim1$ Myr, flux tubes are advected by the moving vortex lines throughout the core, except for a small region around the cusp where the tension force $\vec{f}_B$ is large. The toroidal field deforms into an octupole configuration, which is severely damped by Ohmic diffusion. After $2$ Myr the crustal poloidal field begins to develop an octupole component, due to the magnetic pressures and tensions communicated from the base of the crust through Ohmic diffusion. This is clearly evident at $4$ Myr in Figure~\ref{n_vortex1}. Throughout this simulation Hall drift does not play a major role in the redistribution of the magnetic field, because the field strength is weak, and Ohmic diffusion is the dominant effect ($t_\text{ohm}<t_\text{hall}$). The spin period after $4$ Myr in this simulation is $0.85$ s.

In Model D2 the initial field has strength $\sim10^{13}$ G at the surface, and $1.4\times10^{13}$ G in the core. We set the initial ($300$ yr after neutron star birth) spin period accordingly to $21.8$ ms, and show the results of this simulation in Figure~\ref{n_vortex2}. In the first $5$~kyr vortices cut through flux tubes throughout the core, except for a thin cylinder around the axis of rotation where the vortices move slowly. In the cut-through regime the flux tubes are allowed to bend, and as a result they curve away from the axis of rotation due to the collective drag of outward moving vortices cutting through them. After $10$~kyr the vortices are moving slowly enough that they advect flux tubes throughout the core, except for a thin layer beneath the crust where the flux tube tension is large. In this thin layer the vortices cut through, and the terminal velocity of flux tubes gets very small due to the high density of flux tubes. The sharp curvature of flux tubes at the crust-core interface results in the development of a strong current sheet in the deep crust, which is site to enhanced Ohmic dissipation, and the development of a quadrupole toroidal field through Hall drift.

\begin{figure*}
\centering
\subfloat{\includegraphics[width=1.\textwidth]{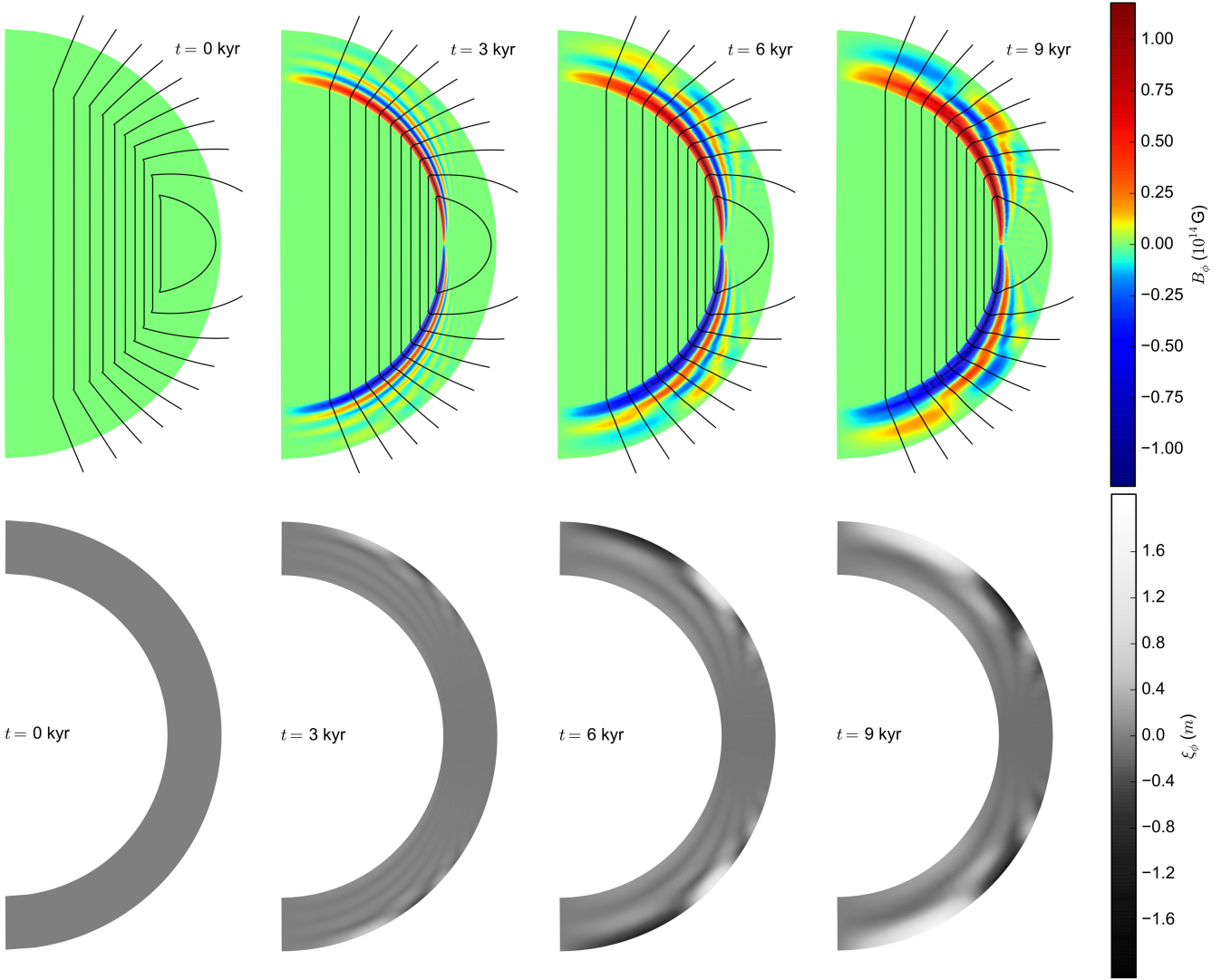}}
\caption{Snapshots of the magnetic field evolution for Model E (Table~\ref{results_table}), shown at $t=0$ kyr, $3$ kyr, $6$ kyr, and $9$ kyr. Top row: The plotting scheme is the same as Figure \ref{HME_core}. Bottom row: Lagrangian displacement of the crust is shown on the colour scale which varies logarithmically, with a linear region around zero. The thickness of the crust has been magnified by a factor or 2.5 in both rows.}
\label{hall_waves}
\end{figure*} 

As the star spins down the number density of vortices decreases along with the critical force $n_v\tilde{f}_v$. As a result, after $\sim100$ kyr the region of cut-through in the outer core begins to grow, allowing the flux tubes to drift once again back into the core, with the Jones drift velocity $\vec{v}_\text{J}$. The thickness of the cut-through layer increases as the star continues to spin down. After $3$ Myr the toroidal field is significantly damped by Ohmic diffusion, and the flux tubes in the core remain curved outward in the outer regions of the core, a result of the collective drag by the vortex lines cutting through the flux tubes. The spin period after $3$ Myr in this simulation is $1.87$ s. 

In Model D3 we use an initial field of strength $\sim2\times10^{13}$ G at the surface, and $\sim2.7\times10^{13}$ G in the core. This field is stronger than previous models, so the star spins down faster, and we set the initial ($300$ yr after neutron star birth) spin period accordingly to $43.5$ ms. The results of this simulation are shown in Figure~\ref{n_vortex3}. In the beginning the vortices cut through in the entire core, except for a thin cylinder around the spin-axis where they are slowly moving. As a result of the drag from the cutting-through vortices, the flux tubes bend away from the axis of rotation, and bunch in the outer core. The sharp curvature of poloidal field lines at the crust-core interface is supported by a strong toroidal current sheet. A toroidal field with quadrupole structure grows in the deep crust, which is site to enhanced Ohmic dissipation. As the star spins down the force of vortices pushing on flux tubes becomes smaller while the tension force grows larger. This continues until $\sim250$ kyr, when the flux tubes stop moving away from the spin-axis, and begin moving back toward. From this point on, the core mostly operates in the cut-through regime, and the combination of the slow spin period, and strong magnetic field means that the flux cannot be expelled from the core. The flux tubes remain bent away from the spin-axis due to the drag of cutting-through vortices for the remainder of the simulation, while the crustal field decays primarily due to Ohmic diffusion. The spin period after $3$~Myr is $4.01$~s. The spin periods we observe in this simulations are not unlike the spin periods of known low-B magnetars. However, it seems unlikely that the toroidal field in these models is strong enough to break the crust and power classical magnetar activity.

\subsection{Elastic back-reaction}\label{magneto-elastic}

In this section we present the coupled evolution of the elastic crustal deformation, and the magnetic field under Hall drift and Ohmic diffusion. The initial magnetic field is chosen such that there is a sharp cusp in the field on the crust-core interface. \cite{goldreich_magnetic_1992} showed that such a disturbance will launch circularly polarized ``Hall waves",  which can propagate from the crust-core interface, and transport magnetic energy toward the surface of a neutron star. \cite{beloborodov_thermoplastic_2014} showed that Hall waves in strong magnetic fields can trigger a thermoplastic instability in the crust, which can generate X-ray activity associated with magnetars. The elastic deformation of the crust can be significant in the upper layers, where the magnetic energy density $\mu_\text{B} = B_z^2/8\pi$ is comparable to the crustal shear modulus $\mu$. Here the crust cannot balance arbitrary stresses generated by Hall drift, so it yields, thus nullifying the Hall effect. In their 1D plane parallel model, \cite{cumming_magnetic_2004} show that the Hall term in the Hall-elastic evolution equation is suppressed by a factor $(1 + \mu_\text{B}/\mu)^{-1}$, so that when $\mu_\text{B}\gg\mu$, Hall drift is significantly suppressed. Unfortunately in this regime ($\mu_\text{B}\gg\mu$), we encounter severe numerical instabilities due to our explicit time integrator. Thus, for now we are restricted to work in the limit $\mu_\text{B}\leq\mu$, where we may still demonstrate the effectiveness of the relaxation method outlined in Section \ref{crust_evolution}.

In Model E we chose the initial field to be purely poloidal, with vertical field lines (pure $B_z$) in the core (in particular $\Psi\propto r_\perp^2$), and a dipole potential field in the crust. The initial field has strength $B\approx 2 \times 10^{14}$ G. There are several physical processes which could cause such a cusp at the crust-core interface in a highly magnetized neutron star, and these motivate our choice of initial field. As demonstrated in Section~\ref{vortex_expulsion}, if the magnetic field is sufficiently weak, superfluid neutron vortices will be present for a significant period during the spin down of a rapidly rotating neutron star, while the ratio of rotational to magnetic energy is high. Transport of flux tubes by outward moving vortices can result in a cusp in the field at the base of the crust, though this will not result in vertical field lines in the core as shown above. Jones flux tube drift in a young magnetar can result in a cusp in the field, and could launch Hall waves, depending on the composition of the core and the subsequent value of the drag coefficient $\alpha$. The launching of short wavelength Hall waves depends on Jones drift being significantly faster than the Hall timescale. There may be other effects which could drive a fast change of the core magnetic field, and thus launch the Hall waves. \cite{beloborodov_magnetar_2016} found that in young magnetars with hot cores ($T_\text{core}\approx10^{9}$ K) and ultra-strong magnetic fields ($B\gtrsim10^{16}$ G), ambipolar diffusion operates in the friction dominated regime, and may cause a fast rearrangment of the core magnetic field. Additionally there may by hydromagnetic instabilities in young magnetars which can result in a rapid rearrangement of the core magnetic field. 

The magneto-elastic evolution in Model E is shown in Figure \ref{hall_waves}. Initially, the cusp in the poloidal field generates strong toroidal currents which in turn generate toroidal field. The result is a burst of Hall waves which propagate away from the core. These waves are the 2D analogue of the Hall waves shown in \cite{li_magnetar_2016}. At 3 kyr the small amplitude, short wavelength Hall waves, have traveled the furthest toward the stellar surface. The long wavelength Hall waves near the core evolve much more slowly. This can be understood if we consider our system as a constant background field, with an oscillating perturbation which is linear in the field. This is valid for the small amplitude Hall waves early in the evolution, which are sufficiently weak such that evolution equation for the poloidal field is weakly coupled to the toroidal field.  This means the structure of the poloidal field is effectively constant. Here we understand the background field to be poloidal, and the perturbation is the Hall waves, early in the evolution before non-linearity becomes significant. The dispersion relation is
\begin{equation}
\omega = \frac{ck|\vec{k}\cdot\vec{B_0}|}{4\pi n_e},
\end{equation}
\citep{goldreich_magnetic_1992} where $\vec{k}$ is the wave vector, $k:=|\vec{k}|$, and $B_0$ a uniform background magnetic field. The waves are seen to fan out from the crust-core interface over the next few kyr, traveling furthest near the poles. This is due to the geometry of the existing background field. The group velocity of Hall waves in the linear regime follows from the dispersion relation \citep{goldreich_magnetic_1992} as 
\begin{equation}
v_{\text{gp}} = \pm \frac{ck[\vec{B_0 + (\vec{\hat{k}}\cdot\vec{B_0})\vec{\hat{k}}}]}{4\pi n_ee},
\end{equation}
where $\vec{\hat{k}}:=\vec{k}/k$. Near the poles, the background field is almost pure $B_r$, so the waves travel radially there. However, near the equator, the background field is almost entirely $B_\theta$, so there is less radial propagation. After the first 6 kyr the waves begin to evolve non-linearly. Hall drift sets in and advects the wave fronts toward the equator of the star. But due to the gradient in electron density, and the fanning of the wave fronts, they are advected non-uniformly, and start to break apart. Diffusion also smears the wave fronts, and decreases the amplitude. The evolution of the crustal displacement is shown in lower panel of Figure~\ref{hall_waves}. The displacement is largest near the surface, where the shear modulus is smallest and the crust yields easily. The crustal displacement reaches a maximum amplitude of $\sim2$ m. 
\section{Discussion}\label{discussion}
In this paper we have modeled the coupled magnetic field evolution of neutron stars in the crust and the core. In the crust we include evolution due to Hall drift and Ohmic diffusion \citep{goldreich_magnetic_1992}, as well as the elastic response of the solid crust. We enforce the correct hydromagnetic equilibrium in the fluid core. We also explore the effects of the Jones flux tube drift \citep{jones_type_2006}, and expulsion by superfluid neutron vortices during spin-down [\cite{ruderman_rotating_1974}, \cite{srinivasan_novel_1990}, \cite{ruderman_neutron_1998}]. In this section we discuss the preliminary implications of these results, in the context of the galactic population of neutron stars, and their observable behavior. 

In Section~\ref{hydromagnetic} we modeled the evolution of an initial poloidal field with broken equatorial symmetry. We evolved the crustal magnetic field through Hall drift and Ohmic diffusion, while the core field evolved according to the hydromagnetic equilibrium we formulate in Section~\ref{hydromagnetic_theory}. We confirm the Hall attractor of \cite{gourgouliatos_hall_2014} for B-fields which penetrate the core, while satisfying the correct hydromagnetic equilibrium. 

In Section~\ref{JonesDrift}, we presented simulations of Jones flux tube drift, which show that the B-field in the core can straighten under the enhanced self tension possessed by the quantized flux tubes. The straightening of flux tubes is associated with the dissipation of free energy stored in the curvature of the field. This straightening occurs on a timescale
\begin{equation}
t_\text{diss} \sim 450 \left(\frac{n_e}{3.5\times 10^{37} \text{ cm}^{-3}} \right)^2\left(\frac{10^{29}\text{ s}^{-1}}{\tilde{\sigma}} \right) \text{ kyr},
\end{equation}
but can occur significantly faster depending on the value of $\tilde{\sigma}$. Interestingly, this can generate a burst of activity in highly magnetized neutron stars which were previously in the Hall attractor state. Importantly, we show that for the range of values of $\tilde{\sigma}$ estimated by \cite{jones_type_2006}, $t_\text{diss}$ is always much smaller than the modified Ohmic timescale $\tilde{t}_\text{ohm}$ [Equation \eqref{modified_ohmic_time}], so that the Ohmic timescale governs the rate of depletion of the global magnetic field. However this timescale is greater by an order of magnitude than the Ohmic timescale of \cite{goldreich_magnetic_1992}.

The timescale for depleting the pulsar magnetic fields in these simulations is very sensitive to the choice of electrical conductivity, and it is worthwhile to discuss the implications of this. Phonon scattering, and impurity scattering are the main ways currents can be diffused in a neutron star crust. Phonon scattering is exponentially suppressed when  $T<T_U = 8.7\times10^{7} $ K $\rho_{14}(Y_e/0.05)(Z/30)^{1/3}$\citep{cumming_magnetic_2004}, and the Umklapp processes freeze out. Impurity scattering is dominant at low temperatures ($T<T_U$), or high impurity levels. Estimates of the impurity levels in the deep crust range from $Q_\text{imp}\approx10^{-3}$ \citep{flowers_evolution_1977}, to $Q_\text{imp}\approx10$ \citep{jones_first-principles_2001}.
For young or accreting pulsars, with temperatures $T\gtrsim10^8$ K, $T>T_U$, and typical impurity levels, phonon scattering will be dominant in the deep crust. Then the timescale for magnetic diffusion in the crust, using the electrical conductivity in Section \ref{crust_evolution} (phonon scattering at $T\approx 2\times10^8$ K), is given by the modified Ohmic timescale \eqref{modified_ohmic_time} $\tilde{t}_\text{ohm}\sim 150$ Myr. A full treatment would include the effects of accretion onto the neutron star surface, and burial of the magnetic field. The consequences of this are not clear, but it should be noted that burial of the field (see e.g. \cite{choudhuri_diamagnetic_2002} and the field configurations therein) could result in suppression of the ratio $B_z/B_x$, leading to an even shorter timescale for the depletion of the global field. \cite{pons_magneto--thermal_2009} found that Ohmic dissipation proceeds faster when thermal feedback on the crustal conductivity is included. This could further shorten the timescale of $150$~Myr we observe in our simulations. 

After a young pulsar has cooled, or accretion has subsided, the neutron star crust will cool. For $T\sim 10^6$ K, $T<T_U$, and impurity scattering will dominate. Then \cite{cumming_magnetic_2004} give the electrical conductivity in the deep crust as
\begin{equation}
\sigma_Q = 4.4\times10^{25} \text{ s}^{-1}(\rho_{14}^{1/3}/Q_\text{imp})(Y_e/0.05)^{1/3}(Z/30).
\end{equation}
In the impurity dominated regime, the timescale for flux to diffuse through the crust is
\begin{equation}
\tilde{t}_\text{ohm} \sim hl \frac{4\pi\sigma_Q}{c^2}\frac{B_z}{B_x}=  \frac{1.8 \text{ Gyr}}{Q_\text{imp}} , 
\end{equation}
meaning that flux is effectively frozen into the crust, and the dipole surface field of the pulsar will no longer decay. For any impurity parameter which yields a decay time comparable to the Hubble time ($Q_\text{imp}\lesssim 0.13$), the field will be approximately stable. This could explain the persistence of magnetic fields in millisecond pulsars, after periods of rapid depletion at higher temperatures. Alternatively, if impurity levels are much higher, as suggested by \cite{jones_first-principles_2001}, then Ohmic diffusion can proceed rapidly in pulsars even after cooling, so that the crust cannot prevent decay of the  dipole field. This would suggest that something elsewhere in the core was inhibiting the motion of flux. 

In the core flux tubes may get caught on magnetized neutron vortices, and be forced to move outward at the same rate. The vortices move outward on a timescale equal to the spin-down time of the star, which is very long for millisecond pulsars. \cite{jones_type_2006} also found that while the outer core is likely a type-II superconductor, protons in the inner core may be type-I. In type-I superconductors magnetic flux is confined to macroscopic filaments of normal matter. Due to the presence of muons in the inner core in some equations of state,  motion of the filaments would be accompanied by the formation of large gradients in chemical potential, limiting the motion of flux to the rate set by weak nuclear interactions \citep{jones_type_2006}. Additionally our simulations do not include a smooth transition from the solid crust to the liquid core. It is possible that some flux tubes get pinned in the pasta phases at the crust core interface, thus causing a remnant field to be left behind. Any of these could provide an explanation for the persistence of a magnetic field in millisecond pulsars despite the decay time we calculate.

In Section~\ref{vortex_expulsion} we modeled the expulsion of flux from the core by the outward motion of neutron vortices during spin-down. We chose the initial spin period for our models by assuming the star was born with a $1$ ms spin period, and allowed to spin down for $300$ years before the phase transition to superfluidity, as suggested by the Cas A remnant [\cite{shternin_cooling_2011}, \cite{page_rapid_2011}]. Models D1, D2, and D3 have typical magnetic field strengths of $5\times10^{12}$ G, $10^{13}$ G, and $2\times10^{13}$ G respectively. 

While \cite{ruderman_neutron_1998} argues that flux tube tension is small compared to the critical force $n_v\tilde{f}_v$, we find that it plays a crucial role in rearranging flux tubes in the core --- even for weaker magnetic fields $\lesssim 5\times10^{12}$ G. This is because the transport of flux tubes by vortices results in the formation of sharp magnetic features, which possess enormous tension, particularly in the outer core, where flux tubes are anchored to the crust. Even in regions where the tension force $\vec{f}_B$ is small, it causes the flux tubes to slide along neutron vortices, and plays an important role in the large scale distribution of flux. We found that when $B\gtrsim2\times10^{13}$ G, the combination of the strong magnetic field and the slower spin period, means that the magnetic field could not be expelled from the core. On the other hand, we found that for $B\lesssim10^{13}$ G, the outward motion of vortices resulted in a partial expulsion of the core magnetic field, into the outer core and deep crust. We find that in all simulations, as the field is pushed away from the spin-axis a toroidal field grows in the deep crust.

When the flux is expelled into the outer core regions, a strong toroidal current sheet develops in the deep crust. These currents drive Ohmic dissipation at an enhanced rate, as compared to core-penetrating fields which vary on larger spatial scales. Additionally, the bunching of flux tubes in the outer core means the poloidal field can be an order of magnitude stronger there, compared to the spin-down inferred dipole field strength. At some stages in our simulations the field configurations loosely resemble the crust-confined fields of \cite{pons_magnetic_2007}, so we may expect thermal emission similar to that in their crust-confined models. The crucial difference is that in our simulations the field penetrates the core. It seems unlikely that flux expulsion could power the magnetar activity of weak-field magnetars or high-B pulsars since the toroidal field in the crust is always $<10^{14}$ G. However, it is possible that flux expulsion could power thermal emission in isolated neutron stars, due to enhanced Ohmic dissipation in the deep crust.

Recent observations of cyclotron emission from the weak-field magnetars SGR 0418+5729 and SWIFT J1822.3-1606 suggest the presence of small-scale magnetic loops near the stellar surface, which can be up to two orders of magnitude stronger than the spin-down inferred dipole field strength [\cite{tiengo_variable_2013}, \cite{rodriguez_castillo_outburst_2016}]. Such strong high-order magnetic multipoles may drive Hall drift on short timescales, and produce X-ray activity normally associated with classical magnetars.

The main shortcoming of our work is that we have not resolved the controversy in the literature between the timescales of \cite{jones_type_2006} and \cite{glampedakis_magnetohydrodynamics_2011}, and this is left for future work. We used the drift timescales derived by Jones, as they lead to interesting dynamical effects at the crust-core interface that are well-modeled in our numerical experiments. Future papers in this series will include the effects of field burial by accretion, and a study of the galactic population of pulsars.

\section*{Acknowledgements}

We thank Daniel Price, Chris Matzner, and Xinyu Li for enlightening discussions. We also thank Kostas Gourgouliatos and Andrew Cumming for helpful correspondence, and providing data files to test our code. We thank the referee for useful comments on the manuscript. This research was supported by an Australian Postgraduate Award (AJB), a Monash Research Acceleration grant (YL), by NASA grants NNX17AK37G, SAO GO6-17063X, SAO GO7-18053X (AMB), and by a grant from the Simons Foundation ($\#$446228, AMB).

%%%%%%%%%%%%%%%%%%%%%%%%%%%%%%%%%%%%%%%%%%%%%%%%%%


\bibliographystyle{mnras}
\bibliography{Astrophysics} % if your bibtex file is called example.bib

%\begin{thebibliography}{99}
%\end{thebibliography}

%%%%%%%%%%%%%%%%%%%%%%%%%%%%%%%%%%%%%%%%%%%%%%%%%%

%%%%%%%%%%%%%%%%% APPENDICES %%%%%%%%%%%%%%%%%%%%%
\appendix

\section{Twist Evolution of the Core.} \label{AppendixA}
 The Hall evolution of $B_\phi$ in the crust can be written as
\begin{equation}
\frac{\partial B_{\phi}}{\partial t} = -\nabla_{\text{p}}\cdot(B_{\phi}\vec{v}_\text{p}) + (r_\bot\vec{B_{\text{p}}}\cdot\nabla_{\text{p}})\left(\frac{v_{\phi}}{r_\bot}\right),
\label{Bphiev}
\end{equation}
with $\vec{v}_p$ and $v_\phi$ the poloidal and toroidal parts of the Hall drift velocity, and we have defined the poloidal differential operator 
\begin{equation}
\nabla_p \equiv \left( \frac{\partial}{\partial r_\bot}, \frac{
\partial}{\partial z}\right),
\end{equation}
using cylindrical coordinates ($r_\bot, z$). The first term on the RHS of \eqref{Bphiev} represents advection of $B_\phi$ by poloidal velocities, and the second term represents shearing of poloidal field lines in the azimuthal direction. By using a combination of the product rule and the divergence constraint, \eqref{Bphiev} can be written in conservative form as
\begin{equation}
\frac{\partial B_{\phi}}{\partial t} + \nabla_\text{p}\cdot\vec{F_\text{hall}}=0,
\end{equation}
where we identify the Hall advection flux 
\begin{equation}
\vec{F_\text{hall}} = B_\phi\vec{v}_\text{p} - v_\phi\vec{B}_\text{p}.
\end{equation}
It is convenient to work in the so-called flux-coordinates $(\Psi,\lambda,\phi)$, where $\Psi$ labels surfaces of constant poloidal flux, and $\lambda$ is the length along a given poloidal field line in the $\phi=const$ plane (see eg. \cite{goedbloed_advanced_2010}). At the base of the crust the boundary condition is $f_\phi = \vec{j}_\text{p}\times\vec{B}_\text{p} /c = 0$, which implies $\vec{v}_\text{p}\parallel\vec{B}_\text{p}$. So the Hall flux can be written in flux coordinates as
\begin{equation}
\vec{F}_{\text{Hall}} = B_\phi|\vec{v}_\text{p}|\hat{e}_{\lambda} - v_\phi|\vec{B}_\text{p}|\hat{e}_{\lambda} = (B_\phi v_\lambda - v_\phi B_\lambda)\hat{e}_{\lambda}=F_\lambda\hat{e}_\lambda,
\end{equation}
where $\vec{B}_\text{p} = B_\lambda\hat{e}_\lambda$, and $\vec{v}_\text{p} = v_\lambda\hat{e}_\lambda$ at the base of the crust. Then, using the scale factors for flux-coordinates
\begin{equation}
h_\psi = \frac{1}{r_\bot B_\lambda},\text{  } h_\lambda = 1,
\end{equation}
we may write the conservation equation for $B_\phi$ in flux coordinates as,
\begin{equation}
\frac{\partial B_\phi}{\partial t} = -r_\bot B_\lambda\frac{\partial}{\partial \lambda}\left(\frac{F_\lambda}{r_\bot B_\lambda}\right).
\end{equation}
Rearranging and integrating both sides with respect to $\lambda$ yields an evolution equation for the twist of the core magnetic field
\begin{equation}
\frac{\partial\zeta(\Psi)}{\partial t} = -[J(\Psi,\lambda_2) - J(\Psi,\lambda_1)],
\end{equation}
where we have identified the twist angle 
\begin{equation}
\zeta(\Psi) = \int_{\lambda_1}^{\lambda_2} d\lambda \left(\frac{B_\phi}{r_\bot B_\lambda}\right),
\label{twistint}
\end{equation}
and the ``flux of twist" into/out of the core as
\begin{equation}
J = \frac{F_\lambda}{r_\bot B_\lambda} = \frac{v_\lambda}{r_\bot}\frac{B_\phi}{B_\lambda} - \frac{v_\phi}{r_\bot}.
\end{equation}

\section{Numerical Details of the Code} \label{AppendixB}
\begin{table}
\caption{The grid resolution used in the crust and core for each of the Models A-E. }
\label{tab:example}
\begin{tabular}{lcccc}
\hline
Model & Crust $(N_r \times N_u)$ & Core $(N_r \times N_u)$ \\
\hline
A & ($100\times133$) & ($400\times133$) \\
B & ($100\times101$) & ($400\times101$) \\
C & ($100\times201$) & --- \\
D1 & ($100\times101$) & ($900\times101$) \\
D2 & ($100\times101$) & ($900\times101$) \\
D3 & ($100\times101$) & ($900\times101$) \\
E & ($500\times201$) & --- \\
\hline
\end{tabular}
\label{grid_table}
\end{table}
We evolve the poloidal and toroidal scalar functions on a discrete grid, which is linear in $r$ and $u\equiv\cos\theta$, in the crust and the core. The variable $u$ varies from -1 at the south pole, to 1 at the north pole, and the radius of the star is $r_*=1$ in units of $10^6$cm. The crust core interface is at $r_c=0.9r_*$. We use the indices $i$ and $j$ to specify grid points in the $r$ and $u$ directions respectively. In most simulations the index $j$ varies from $j_1=-50$ corresponding to the south pole, to $j_2=50$ corresponding to the north pole, with $j=0$ defining the equator. We choose the difference in $u$ such that $\delta u = 2/(j_2-j_1)$. The index $i$ varies from $i_0=0$ at the centre ($r=0$), to typical values of $i_c=400$ at the crust core interface ($r=r_c$) depending on the simulation. Throughout the crust and the last few rows of the core (ghost points for the crust) the radial grid spacing is $\delta r_\text{crust}=1/i_s$. The radial grid spacing in the outer few rows of the core grid matches the radial grid spacing of the crust, for ease of implementing boundary conditions on the crustal field. In order to avoid numerical instabilities near the poles in some simulations, we added adjustable patches of increased resolution in the $u$ direction. Depending on the magnetic field structure, angular resolution was some times set to 3 times the original resolution near the poles in order to obtain convergence. This resolved the issue, and added little expense to the computations. 

We evaluate spatial derivatives on the RHS of the crustal evolution equations [Equations \eqref{hall_pol}, \eqref{hall_tor}, \eqref{elasto_mag}] with the following finite difference formulae. To evaluate the radial derivatives at each time step in the crust we use
\begin{equation}
\Psi_r = \frac{\Psi_{j, i+1} - \Psi_{j, i-1}}{2\delta r},
\end{equation}
\begin{equation}
\Psi_{rr} = \frac{\Psi_{j, i-1} -2\Psi_{j,i}+ \Psi_{j, i+1}}{\delta r^2},
\end{equation}
with the subscript a short hand for partial derivative. For the derivatives with respect to $u$ however, a different approach was needed, since central differences do not preserve second order accuracy on a non-uniform grid. We use the following finite differences which are generalized to maintain second order accuracy [Equations A3b, and A4c in \cite{bowen_derivative_2005}]. These are found by using Lagrange interpolation to fit a polynomial to the points, and then taking a derivative of that polynomial. We first define the displacements $\alpha_j = u_j - u$, where $u$ is the point at which we evaluate the derivative, and $u_j$ is a grid point. The point $u$ may be any point contained by the grid points $(j-1,j,j+1)$, not necessarily a grid point. The first and second derivatives with respect to $u$ are then
\begin{equation}
\begin{split}
\Psi_u =& -\frac{(\alpha_2 + \alpha_3)\Psi_{j-1,i}}{(\alpha_1-\alpha_2)(\alpha_1-\alpha_3)} - \frac{(\alpha_1+\alpha_3)\Psi_{j,i}}{(\alpha_2 - \alpha_1)(\alpha_2 - \alpha_3)}
\\[1mm]&
- \frac{(\alpha_1 + \alpha_2)\Psi_{j+1,i}}{(\alpha_3 - \alpha_1)(\alpha_3 - \alpha_2)},
\end{split}
\label{diff1}
\end{equation}
\begin{equation}
\begin{split}
\Psi_{uu} = &-\frac{2(\alpha_2 + \alpha_3 + \alpha_4)\Psi_{j-2,i}}{(\alpha_1 - \alpha_2)(\alpha_1 - \alpha_3)(\alpha_1 - \alpha_4)} 
\\[1mm]&
- \frac{2(\alpha_1 + \alpha_3 + \alpha_4)\Psi_{j-1,i}}{(\alpha_2 - \alpha_1)(\alpha_2 - \alpha_3)(\alpha_2 - \alpha_4)} 
\\[1mm]&
- \frac{2(\alpha_1 + \alpha_2 + \alpha_4)\Psi_{j,i}}{(\alpha_3 - \alpha_1)(\alpha_3 - \alpha_2)(\alpha_3 - \alpha_4)} 
\\[1mm]&
- \frac{2(\alpha_1 + \alpha_2 + \alpha_3)\Psi_{j+1,i}}{(\alpha_4 - \alpha_1)(\alpha_4 - \alpha_2)(\alpha_4 - \alpha_3)}.
\end{split}
\label{diff2}
\end{equation}
We use the same derivative formula for the toroidal scalar function $I$. 

For evolution equations in the core [Equations \eqref{diffusion_relax}, \eqref{jones_eqn}, \eqref{vortex_eqn}] we use the difference Equations \eqref{diff1} and \eqref{diff2} for derivatives in the $u$ direction. We also require specialized formula for radial derivatives in the core, because the radial grid spacing changes in the outer few rows of the core grid. We use formula the same as those above [Equations \eqref{diff1} and \eqref{diff2}], but with differences in the radial direction ($u\longrightarrow r$, $j\longrightarrow i$). 

We use a variable time step, which shrinks in order to avoid instability in the evolution. Because we are evolving magnetic fields and crustal displacements with a variety of evolution equations, we calculate a stable time step for each evolution equation. For the Hall effect [Equations \eqref{hall_pol} and \eqref{hall_tor}], we use the fastest electron velocity in the grid to limit the maximum time step, using
\begin{equation}
\delta t_\text{hall} = k_c \frac{4\pi n_e e}{c}\frac{\delta r \delta \mu}{|\nabla\times\vec{B}_\text{T}|},
\end{equation}
with $k_c$ a Courant parameter. For Ohmic diffusion [RHS of Equations \eqref{hall_pol} and \eqref{hall_tor}], we use
\begin{equation}
\delta t_\text{ohm} = k_c\frac{\delta r^2}{\eta} = k_c\frac{4\pi\sigma\delta r^2}{c},
\end{equation}
which is minimized by choosing the smallest conductivity. For the elastic relaxation [Equation~\eqref{elasto_mag}] we use the time step
\begin{equation}
\delta t_\text{el} = k_c \gamma\frac{\delta r^2}{v_\text{sh}^2}.
\end{equation}
For the hydromagnetic relaxation Equation~\eqref{diffusion_relax} we choose the diffusion time step
\begin{equation}
\delta t_\text{hme} = k_c\frac{\delta r^2}{k}.
\end{equation}
The stable time step for Jones drift [Equation~\eqref{jones_eqn}] is chosen using the maximum flux tube velocity on the computational grid,
\begin{equation}
\delta t_\text{Jones} = k_c \frac{\delta r \delta\mu }{\text{max}|\vec{v}_\text{J}|},
\end{equation}
and similarly with the spin-down transport of flux [Equation~\eqref{vortex_eqn}]
\begin{equation}
\delta t_\text{sd} = k_c \frac{\delta r \delta\mu }{\text{max}|\vec{v}_\text{sd}|}.
\end{equation}
In all of the above, the Courant parameter $0<k_c<1$ is chosen so that such that we observe convergence and stability. At each time step, we evaluate the RHS of the evolution equations using the difference formulae above, then use Euler integration with the smallest time step
\begin{equation}
\delta t = \text{min}\lbrace dt_\text{hall}, \delta t_\text{ohm}, \delta t_\text{el}, \delta t_\text{hme}, \delta t_\text{Jones}, \delta t_\text{sd}\rbrace,
\end{equation}
to advance the functions $\Psi$ and $I$ to $t+\delta t$. In the code we normalize the Hall Evolution equation in the same way as \cite{gourgouliatos_hall_2014},
\begin{equation}
\begin{split}
1.6\times 10^6 \frac{\partial \vec{B_{14}}}{\partial t_{\text{yr}}} =& -\tilde{\nabla}\times\left(\frac{\tilde{\nabla}\times\vec{B_{14}}}{n_{e,0}}\times\vec{B_{14}}\right)
\\[2mm]&
- 0.02\tilde{\nabla} \times \left( \frac{\tilde{\nabla}\times\vec{B_{14}}}{\sigma_0} \right),
\end{split}
\end{equation}
where $\vec{B_{14}} = \vec{B}/10^{14}$ G, $t_\text{yr} = t/3.15\times 10^7$ $s$, $\tilde{\nabla}$ is the del operator with lengths normalized to $10^6$ cm, $n_{e,0}=n_e/2.5\times10^{34}$ cm$^{-3}$, and $\sigma_0=\sigma/1.8\times10^{23}$ $\text{s}^{-1}$. In Model A we choose the relaxation parameter $k=2\times 10^{-6} \text{ cm s}^{-1}$ In Model E we choose the parameter $\gamma=1\times10^{16}\text{ s}^{-1}$.

In the core we treat coordinate singularities along the pole, and at the origin by freezing in the magnetic field beyond some flux surface $\Psi_0$, so that the magnetic field is unevolving very close to the pole. This is done by multiplying the velocity fields in Equations \eqref{diffusion_relax} \eqref{jones_eqn} and \eqref{vortex_eqn} by the function
\begin{equation}
 s(\Psi) = \frac{1}{\text{exp}[-a(\Psi - \Psi_0)] + 1},
\end{equation}
which behaves like a smoothed step function. The parameter $a$ is chosen to make the step as steep as possible while still being resolved by the grid mesh. In Model A we also add a term to the RHS of Equation \eqref{diffusion_relax},
\begin{equation}
-\frac{1}{\tau}\frac{I(r,\theta)}{\text{exp}[b(\Psi - \Psi_0)] + 1},
\end{equation}
to ensure that any toroidal field beyond $\Psi_0$ is exponentially reduced on the timescale $\tau$. Our results are not sensitive to these methods, so long as $\Psi_0$ is close to the pole.

We have tested the Ohmic evolution of our code by comparing with the analytic Ohmic eigenmodes, and observe excellent agreement. We also study the agreement of our code with the grid based code of \cite{gourgouliatos_hall_2014}, (data files provided by the authors). We compare the Hall-Ohmic evolution of 3 initial fields--the so-called ``Hall Equilibrium", ``Ohmic Eigenmode", and ``Barotropic MHD Equilibrium", shown in Figure~2 of \citep{gourgouliatos_hall_2014}. Excellent agreement was observed in all cases. We have also carried out resolution studies of all simulations presented in Section~\ref{results}, and summarize the grid sizes for which each simulation had converged in Table~\ref{grid_table}. 

%%%%%%%%%%%%%%%%%%%%%%%%%%%%%%%%%%%%%%%%%%%%%%%%%%

% Don't change these lines
\bsp	% typesetting comment
\label{lastpage}
\end{document}